\documentclass[prd,aps,nofootinbib,notitlepage,showpacs,preprintnumbers]
{revtex4-1}
\usepackage{graphicx,epsf,amsmath,amsfonts,amssymb,amsbsy}
\usepackage{epsfig}
\newcommand{\ds}{\displaystyle}
\newcommand{\vev}[1]{\langle#1\rangle}
\newcommand{\mat}{\left ( \begin{array}}
\newcommand{\emat}{\end{array} \right )}
\newcommand{\vect}{\left ( \begin{array}{c}}
\newcommand{\evect}{\end{array} \right )}

\begin{document}

%\hfill HU-EP-11/11

\title{Superconducting phase transitions induced by chemical potential in
(2+1)-dimensional four-fermion quantum field theory}
\author{K.G. Klimenko $^{a,b}$,
R.N. Zhokhov $^{a}$ and V.Ch. Zhukovsky $^{c}$}
\affiliation{$^{a}$ Institute for High Energy Physics,
142281, Protvino, Moscow Region, Russia}
\affiliation{$^{b}$ University "Dubna" (Protvino branch),
142281, Protvino, Moscow Region, Russia}
\affiliation{$^{c}$ Faculty of
Physics, Moscow State University, 119991, Moscow, Russia}
%\email{debert@physik.hu-berlin.de}
%\affiliation{$^{1)}$ Institute of Physics, Humboldt-University
%Berlin, 12489 Berlin, Germany} \affiliation{$^{2)}$ Faculty of
%Physics, Moscow State University, 119991, Moscow, Russia}
%\affiliation{$^{3)}$ IHEP and University "Dubna" (Protvino branch),
%142281, Protvino, Moscow Region, Russia}

\begin{abstract}
In the paper a generalization of the (1+1)-dimensional model by Chodos et al [Phys. Rev. D61, 045011 (2000)] has been performed to the case of
(2+1)-dimensional spacetime. The model includes four-fermion interaction
both in the fermion-antifermion (or chiral) and fermion-fermion (or
superconducting) channels. We study temperature $T$ and chemical potential $\mu$ induced phase transitions in the leading order of large-$N$
expansion technique, where $N$ is a number of fermion fields. It is shown that  at sufficiently large values of $\mu$ and arbitrary relations between
coupling constants, superconducting  phase appears in the system both at
$T=0$ and $T>0$. In particular, at $T=0$ and sufficiently weak
attractive interaction in the chiral channel, the Cooper pairing
occurs for arbitrary couplings in the superconducting channel even at infinitesimal values of $\mu$.
\end{abstract}
%\pacs{12.39.Ki, 12.38.Mh, 21.65.Qr}

%%% 12.38.Mh Quark-gluon plasma
%%% 21.65.Qr Quark matter
%%% 12.39.Ki Relativistic quark model

\maketitle

\section{ Introduction}

Last years great attention has been paid to investigation of
(2+1)-dimensional quantum field theories (QFT) and, in particular,
to models with four-fermion interactions of the Gross-Neveu (GN) \cite{GN} type. Partially, this interest is explained by more simple  structure
of QFT in two-, rather than in three spatial dimensions. As a result, it is
much easier to investigate qualitatively such real physical
phenomena as dynamical symmetry breaking
\cite{GN,semenoff,rosenstein,klimenko2,inagaki,hands,bashir,khanna} and color superconductivity \cite{toki}, and to model phase diagrams of real quantum chromodynamics (QCD) \cite{kneur} etc. in the framework of
(2+1)-dimensional models. Another example of
this kind is spontaneous chiral symmetry breaking induced by external
magnetic or chromomagnetic fields. This effect was for the first time studied  also in terms of (2+1)-dimensional GN model \cite{klimenko}. Moreover, these theories are very useful in developing new QFT techniques like the optimized perturbation theory \cite{kneur,k}, and so on.

However, there is yet another more serious motivation for studying (2+1)-dimensional QFT. It is supported by the fact that there
are many condensed matter systems which,
firstly, have a (quasi-)planar structure and, secondly, their
excitation spectrum is described adequately by relativistic
Dirac-like equation rather than by Schr\"{o}dinger one. Among these
systems are the high-T$_c$ cuprate and iron  superconductors
\cite{davydov}, the one-atom thick layer of carbon atoms, or
graphene, \cite{niemi,castroneto} etc. Thus, many properties of
such condensed matter systems can be explained in the framework of
various (2+1)-dimensional QFT, including the GN-type models (see,
e.g., \cite{Semenoff:1998bk,babaev,gorbar,Fialkovsky,Cortijo,caldas,zkke,ohsaku,marino} and references therein).

In this paper we study phase transitions in a (2+1)-dimensional
GN-type model which describes competition between two processes: chiral symmetry breaking (excitonic pairing) and superconductivity (Cooper pairing). Clearly, the model is suitable for qualitative analysis of
superconducting phase transitions in quasi-planar condensed matter
systems. The structure of our model is a direct generalization of known
(1+1)-dimensional model of Chodos et al. \cite{chodos,abreu}, which remarkably mimics the temperature $T$ and chemical potential $\mu$
phase diagram of real QCD, to the case of (2+1)-dimensional
spacetime. Recall that in \cite{chodos}, in order to avoid the
prohibition on Cooper pairing as well as spontaneous breaking of
continuous symmetry in (1+1)-dimensional models (known as the
Mermin-Wagner-Coleman no-go theorem \cite{coleman}), the
consideration was performed in the leading order of $1/N$-technique,
i.e. in the large-$N$ limit assumption, where $N$ is the number of
fermion fields. In this case quantum fluctuations, which would
otherwise destroy a long-range order corresponding to spontaneous
symmetry breaking, are suppressed by $1/N$ factors. By the same reason in
(2+1)-dimensional spacetime and in the case of finite values of $N$,
spontaneous breaking of continuous symmetry is allowed only at
zero temperature, i.e. it is forbidden at $T>0$. Hence, in order to
make  investigation of superconducting phase transitions possible
at $T>0$, we suppose, as it was done in \cite{chodos}, that in the
framework of our model $N\to\infty$.

So at $T=0$ the results of our paper may be aplied for the description
of superconductivity in different $N$-layer condensed matter systems
($N$ is finite and can even be equal to one), whereas at $T>0$ it is
better to use  the results in the description of macroscopic systems
composed  of a very large number of layers, such as graphite, etc.

The paper is organized as follows. In Sec. II the GN-type model
with four-fermion ineractions in the fermion-antifermion (or chiral) and fermion-fermion (or superconducting) channels is presented. Here the unrenormalized thermodynamic potential (TDP) of the model is obtained in the leading order of large-$N$ expansion technique. In the next Sec. III
a renormalization group invariant expression for the TDP is obtained
whose global minimum point provides us with chiral and Cooper pairs
condensates. In Sec. IV  phase structure of the model is described at
$T=0$ both at $\mu=0$ and $\mu\ne 0$. In particular, it is established in this Section that infinitesimal chemical potential induces the superconductivity phenomenon in the case of a rather weak attractive interaction in the fermion-antifermion channel. Finally, in Sec. V the $(\mu,T)$-phase
diagrams are presented for some representative values of coupling
constants. We show in this Section that at arbitrary fixed $T>0$
superconductivity is induced in the system at sufficiently large
values of $\mu$. Some related problems of our consideration are
relegated to three Appendices.

\section{ The model and its thermodynamic potential}
\label{effaction}

Our investigation is based on a (2+1)-dimensional GN--type model
with massless fermions belonging to a fundamental multiplet of the
auxiliary  $O(N)$ flavor group. Its Lagrangian describes the
interaction both in the scalar fermion--antifermion and scalar
difermion channels:
\begin{eqnarray}
 L=\sum_{k=1}^{N}\bar \psi_k\Big (\gamma^\nu i\partial_\nu
+\mu\gamma^0\Big )\psi_k+ \frac {G_1}N\left (\sum_{k=1}^{N}\bar
\psi_k\psi_k\right )^2+\frac {G_2}N\left (\sum_{k=1}^{N} \psi_k^T
C\psi_k\right )\left (\sum_{j=1}^{N}\bar \psi_jC\bar\psi_j^T\right
), \label{1}
\end{eqnarray}
where $\mu$ is the fermion number chemical potential (see also the
comments after formula (3)). As noted above, all fermion
fields $\psi_k$ ($k=1,...,N$) form a fundamental multiplet of $O(N)$
group. Moreover, each field $\psi_k$ is a four-component
Dirac spinor (the symbol $T$ denotes the transposition operation).
The quantities $\gamma^\nu$ ($\nu =0,1,2$) are matrices in the
4-dimensional spinor space. Moreover, $C\equiv\gamma^2$ is the
charge conjugation matrix. The algebra of the $\gamma^\nu$-matrices
as well as their particular representation are given in Appendix
\ref{ApA}. Clearly, the Lagrangian $L$ is invariant under
transformations from the internal auxiliary $O(N)$ group, which is
introduced here in order to make it possible to perform all the
calculations in the framework of the nonperturbative large-$N$
expansion method. Physically more interesting is that the model (1)
is invariant under the discrete chiral transformation,
$\psi_k\to\gamma^5\psi_k$ (the particular realization of the
$\gamma^5$-matrix is presented in Appendix \ref{ApA}), as well as
with respect to the transformations from the continuous $U(1)$
fermion number group, $\psi_k\to\exp (i\alpha)\psi_k$ ($k=1,...,N$),
responsible for the fermion number conservation or, equivalently,
for the electric charge conservation law in the system under
consideration.

The linearized version of Lagrangian (\ref{1}) that
contains auxiliary  bosonic fields $\sigma (x)$, $\Delta(x)$ and
$\Delta^{*}(x)$ has the following form
\begin{eqnarray}
{\cal L}\ds = -\frac{N\sigma^2}{4G_1} -\frac N{4G_2}\Delta^{*}\Delta+
\sum_{k=1}^{N}\left [\bar\psi_k\Big (\gamma^\nu i\partial_\nu
+\mu\gamma^0 -\sigma \Big )\psi_k-
 \frac{\Delta^{*}}{2}\psi_k^TC\psi_k
-\frac{\Delta}{2}\bar\psi_k C\bar\psi_k^T\right ]. \label{2}
\end{eqnarray}
Clearly, the Lagrangians
(\ref{1}) and (\ref{2}) are equivalent, as can be seen by using the
Euler-Lagrange equations of motion for  bosonic fields which take the
form
\begin{eqnarray}
\sigma (x)=-2\frac {G_1}N\sum_{k=1}^{N}\bar\psi_k\psi_k,~~ \Delta(x)=-2\frac
{G_2}N\sum_{k=1}^{N}\psi_k^TC\psi_k,~~ \Delta^{*}(x)=-2\frac {G_2}N\sum_{k=1}^{N}\bar\psi_k
C\bar\psi_k^T. \label{3}
\end{eqnarray}
One can easily see from (\ref{3}) that the neutral field $\sigma(x)$
is a real quantity, i.e. $(\sigma(x))^\dagger=\sigma(x)$ (the
superscript symbol $\dagger$ denotes the Hermitian conjugation), but
the (charged) difermion fields $\Delta(x)$ and $\Delta^*(x)$ are
mutually Hermitian conjugated complex quantities, so $(\Delta(x))^\dagger=
\Delta^{*}(x)$ and vice versa. Clearly, all the fields (\ref{3}) are
singlets with respect to the auxiliary $O(N)$ group. \footnote{Note
that the $\Delta (x)$ field is a flavor O(N) singlet, since the
representations of this group are real.} Moreover, with respect to
parity transformation $P$ (see also the comment in Appendix A),
\begin{eqnarray}
P~:~~\psi_k (t,x,y)\to\gamma^5\gamma^1\psi_k (t,-x,y),~~~~k=1,...,N,
\label{03}
\end{eqnarray}
the fields $\sigma(x)$, $\Delta (x)$ and $\Delta^{*}(x)$ are even
quantities, i.e. they are scalars. If the difermion field
$\Delta(x)$ has a nonzero ground state expectation value, i.e.
$\vev{\Delta(x)}\ne 0$, the Abelian fermion number $U(1)$ symmetry
of the model is spontaneously broken down and the superconducting
phase is realized in the model. However, if $\vev{\sigma (x)}\ne 0$
then the discrete chiral symmetry of the model is spontaneously
broken.

Let us now study the phase structure of the four-fermion model (1)
starting from the equivalent semi-bosonized Lagrangian (\ref{2}).
In the leading order of the large-$N$ approximation, the effective
action ${\cal S}_{\rm {eff}}(\sigma,\Delta,\Delta^{*})$ of the
considered model is expressed by means of the path integral over
fermion fields
$$
\exp(i {\cal S}_{\rm {eff}}(\sigma,\Delta,\Delta^{*}))=
  \int\prod_{l=1}^{N}[d\bar\psi_l][d\psi_l]\exp\Bigl(i\int {\cal
  L}\,d^3 x\Bigr),
$$
where
\begin{eqnarray}
&&{\cal S}_{\rm {eff}} (\sigma,\Delta,\Delta^{*}) =-\int
d^3x\left [\frac{N}{4G_1}\sigma^2(x)+
\frac{N}{4G_2}\Delta (x)\Delta^{*}(x)\right ]+ \widetilde {\cal
S}_{\rm {eff}}. \label{5}
\end{eqnarray}
The fermion contribution to the effective action, i.e.\  the term
$\widetilde {\cal S}_{\rm {eff}}$ in (\ref{5}), is given by:
\begin{equation}
\exp(i\widetilde {\cal S}_{\rm
{eff}})=\int\prod_{l=1}^{N}[d\bar\psi_l][d\psi_l]\exp\Bigl\{i\int
\sum_{k=1}^{N}\Big
[\bar\psi_k(\gamma^\nu i\partial_\nu +\mu\gamma^0 -\sigma
)\psi_k -
 \frac{\Delta^{*}}{2}\psi_k^T C\psi_k
-\frac{\Delta}{2}\bar\psi_kC \bar\psi_k^T\Big ]d^3 x\Bigr\}.
\label{6}
\end{equation}
The ground state expectation values $\vev{\sigma(x)}$,
$\vev{\Delta(x)}$, and $\vev{\Delta^*(x)}$ of the composite bosonic
fields are determined by the saddle point equations,
\begin{eqnarray}
\frac{\delta {\cal S}_{\rm {eff}}}{\delta\sigma (x)}=0,~~~~~
\frac{\delta {\cal S}_{\rm {eff}}}{\delta\Delta (x)}=0,~~~~~
\frac{\delta {\cal S}_{\rm {eff}}}{\delta\Delta^* (x)}=0. \label{7}
\end{eqnarray}
For simplicity, throughout the paper we suppose that the above
mentioned ground state expectation values do not depend on
space-time coordinates, i.e.
\begin{eqnarray}
\vev{\sigma(x)}\equiv M,~~~\vev{\Delta(x)}\equiv \Delta,~~~
\vev{\Delta^*(x)}\equiv \Delta^*, \label{8}
\end{eqnarray}
where $M,\Delta,\Delta^*$ are constant quantities. In fact, they are
coordinates of the global minimum point of the thermodynamic
potential (TDP) $\Omega (M,\Delta,\Delta^*)$.
 In the leading order of the large-$N$ expansion it is defined by the
 following expression:
\begin{equation*}
\int d^3x \Omega (M,\Delta,\Delta^*)=-\frac{1}{N}{\cal S}_{\rm
{eff}}\{\sigma(x),\Delta (x),\Delta^*(x)\}\Big|_{\sigma
    (x)=M,\Delta(x)= \Delta,\Delta^*(x)=\Delta^*} ,
\end{equation*}
which gives
\begin{eqnarray}
\int d^3x\Omega (M,\Delta,\Delta^*)\,\,&=&\,\,\int d^3x\left
(\frac{M^2}{4G_1}+\frac{\Delta\Delta^*}{4G_2}\right
)+\frac{i}{N}\ln\left (
\int\prod_{l=1}^{N}[d\bar\psi_l][d\psi_l]\exp\Big (i\int\sum_{k=1}^{N}\Big
[\bar\psi_k D\psi_k\right.\nonumber\\&& \left.-
\frac{\Delta^*}{2}\psi_k^TC\psi_k -\frac{\Delta}{2}\bar\psi_k
C\bar\psi_k^T\Big ] d^3 x \Big )\right ), \label{9}
\end{eqnarray}
where $D=\gamma^\nu i\partial_\nu +\mu\gamma^0-M$. To proceed, let
us first point out that without loss of generality the quantities
$\Delta,\Delta^*$ might be considered as real ones. \footnote{Otherwise,  phases of the complex values $\Delta,\Delta^*$
might be eliminated by an appropriate transformation of fermion
fields in the path integral (\ref{9}).} So, in the following we will
suppose that $\Delta=\Delta^*\equiv\Delta$, where $\Delta$ is
already a real quantity. Then, in order to find a convenient
expression for the TDP it is necessary to invoke Appendix B, where
the path integral similar to (\ref{9}) is evaluated. \footnote{In
Appendix \ref{ApB} we consider for simplicity the case $N=1$,
however the procedure is easily generalized to the case with $N>1$.}
So, taking into account in (\ref{9}) the relation (\ref{B8}) we
obtain the following expression for the zero temperature,
$T=0$, TDP of the GN model (1):
\begin{eqnarray}
\Omega (M,\Delta)=
\frac{M^2}{4G_1}+\frac{\Delta^2}{4G_2}%\nonnumber\\
+i\int\frac{d^3p}{(2\pi)^3}\ln\Big [(p_0^2-({\cal
E}_\Delta^+)^2)(p_0^2 -({\cal E}_\Delta^-)^2)\Big ],
\label{12}
\end{eqnarray}
where $({\cal E}_\Delta^\pm)^2=E^2+\mu^2+\Delta^2\pm 2
\sqrt{M^2\Delta^2+\mu^2E^2}$ and $E=\sqrt{M^2+|\vec p|^2}$.
%%%%%%%%%%%%
 Obviously, the
function $\Omega (M,\Delta)$ is invariant under
each of the transformations $M\to-M$, $\Delta\to -\Delta$ and $\mu\to-\mu$.  Hence, without loss of generality, we restrict
ourselves to the constraints $M\ge 0$, $\Delta\ge 0$ and $\mu\ge 0$ and  will investigate the global
minimum point of the TDP (\ref{12}) just on this region.
%%%%%%%%%%%%%%%
%Throughout the paper we suppose that $\mu\ge 0$, $M\ge 0$ and
%$\Delta\ge 0$.
Using in the expression (\ref{12}) a rather general formula
\begin{eqnarray}
\int_{-\infty}^\infty dp_0\ln\big
(p_0-A)=\mathrm{i}\pi|A|,\label{int}
\end{eqnarray}
where $A$ is a real quantity, it is possible to reduce it to the
following one:
\begin{eqnarray}
\Omega (M,\Delta)\equiv\Omega^{un} (M,\Delta)=
\frac{M^2}{4G_1}+\frac{\Delta^2}{4G_2}%\nonnumber\\
-\int\frac{d^2p}{(2\pi)^2}\Big ({\cal E}_\Delta^++{\cal
E}_\Delta^-\Big ).  \label{13}
\end{eqnarray}
The integral term in (\ref{13}) is an ultraviolet divergent one,
hence to obtain any information from this expression we need to
renormalize it.

\section{The renormalization procedure at $T=0$}

First of all, let us regularize the zero temperature TDP (\ref{13})
by cutting momenta, i.e. we suppose that $|p_1|<\Lambda$,
$|p_2|<\Lambda$ in (\ref{13}). As a result we have the following
regularized expression (which is finite at finite values of
$\Lambda$):
\begin{eqnarray}
\Omega^{reg} (M,\Delta)=
\frac{M^2}{4G_1}+\frac{\Delta^2}{4G_2}%\nonnumber\\
-\frac{1}{\pi^2}\int_0^\Lambda dp_1\int_0^\Lambda dp_2\Big ({\cal E}_\Delta^++{\cal E}_\Delta^-\Big ). \label{15}
\end{eqnarray}
Let us use in (\ref{15}) the following asymptotic expansion
\begin{eqnarray}
{\cal E}_\Delta^++{\cal E}_\Delta^-=2|\vec p|+\frac{M^2+\Delta^2}{|\vec p|}+{\cal O}(1/|\vec p|^3),\label{16}
\end{eqnarray}
where $|\vec p|=\sqrt{p_1^2+p_2^2}$. Then, upon integration there
term-by-term, it is possible to find
 \begin{eqnarray}
\Omega^{reg}(M,\Delta)&=&M^2\left [\frac
1{4G_1}-\frac{2\Lambda\ln(1+\sqrt{2})}{\pi^2}\right
]\nonumber\\&+&\Delta^2\left [\frac
1{4G_2}-\frac{2\Lambda\ln(1+\sqrt{2})}{\pi^2}\right
]-\frac{2\Lambda^3(\sqrt{2}+\ln(1+\sqrt{2}))}{3\pi^2}+{\cal
O}(\Lambda^0), \label{17}
\end{eqnarray}
where ${\cal O}(\Lambda^0)$ denotes an expression which is finite in
the limit $\Lambda\to \infty$.  Second, we suppose that the bare
coupling constants $G_1$ and $G_2$ depends on the cutoff parameter
$\Lambda$ in such a way that in the limit
$\Lambda\to\infty$ one obtains a finite expressions in the square
brackets of (\ref{17}). Clearly, to fulfil this requirement it is
sufficient to require that
 \begin{eqnarray}
\frac 1{4G_1}\equiv \frac
1{4G_1(\Lambda)}=\frac{2\Lambda\ln(1+\sqrt{2})}{\pi^2}+\frac{1}{2\pi g_1},
~~~\frac 1{4G_2}\equiv \frac
1{4G_2(\Lambda)}=\frac{2\Lambda\ln(1+\sqrt{2})}{\pi^2}+\frac{1}{2\pi g_2},
\label{18}
\end{eqnarray}
where $g_{1,2}$ are finite and $\Lambda$-independent model parameters
with dimensionality of inverse mass.
%%%%%%%%
Moreover, since bare couplings $G_1$ and $G_2$
%are renormalization group invariant quantities,
do not depend on a normalization point, the same
property is also valid for $g_{1,2}$.
%%%%%%%%%%
Hence, taking
into account in (\ref{15}) and (\ref{17}) the relations (\ref{18}) and
ignoring there an infinite $M$- and $\Delta$-independent constant,
one obtains the following {\it renormalized}, i.e. finite, expression
for the TDP
\begin{eqnarray}
\Omega^{ren}(M,\Delta)&=&\lim_{\Lambda\to\infty}
\left\{\Omega^{reg}(M,\Delta)\Big |_{G_1= G_1(\Lambda),G_2=
G_2(\Lambda)}+\frac{2\Lambda^3(\sqrt{2}+\ln(1+\sqrt{2}))}{3\pi^2}\right\}.\label{19}
\end{eqnarray}
It should also be mentioned that the TDP (\ref{19}) is a
renormalization group invariant quantity.

The fact that it is possible to renormalize the effective potential
of the initial model (1) in the leading order of the large
$N$-expansion is the reflection of a more general property of
(2+1)-dimensional theories with four-fermion interactions. Indeed,
it is well known that in the framework of  the ''naive''
perturbation theory (over coupling constants) these models are not
renormalizable. However, as it was proved in \cite{rosenstein},
in the framework of nonperturbative large $N$-technique these models
are renormalizable in each order of $1/N$-expansion.

In vacuum, i.e. at $\mu=0$, the ${\cal O}(\Lambda^0)$ term in
(\ref{17}) can be calculated explicitly, so we have for the
renormalized effective potential $V(M,\Delta)$ the
expression \footnote{Vacuum TDP is usually called effective potential.}
\begin{eqnarray}
V(M,\Delta)\equiv \Omega^{ren}(M,\Delta)\big |_{\mu=0}=
\frac{M^2}{2\pi g_1}+\frac{\Delta^2}{2\pi g_2}+\frac{(M+\Delta)^{3}}{6\pi}+\frac{|M-\Delta|^{3}}{6\pi}.\label{25}
\end{eqnarray}

Now, let us obtain an alternative expression for the renormalized TDP
(\ref{19}) at $\mu\ne 0$. For this purpose one can rewrite the
unrenormalized TDP $\Omega^{un}(M,\Delta)$ (\ref{13}) in the following
way
\begin{eqnarray}
\Omega^{un} (M,\Delta)&=&
\frac{M^2}{4G_1}+\frac{\Delta^2}{4G_2}-\int\frac{d^2p}{(2\pi)^2}\left ({\cal E}_\Delta^+\big |_{\mu=0}+{\cal E}_\Delta^-\big |_{\mu=0}\right )
-\int\frac{d^2p}{(2\pi)^2}\Big ({\cal E}_\Delta^++{\cal E}_\Delta^--{\cal E}_\Delta^+\big |_{\mu=0}-{\cal E}_\Delta^-\big |_{\mu=0}\Big ), \label{013}
\end{eqnarray}
where
 \begin{eqnarray*}
{\cal E}_\Delta^+\big |_{\mu=0}+{\cal E}_\Delta^-\big |_{\mu=0}=
\sqrt{|\vec p|^2+(M+\Delta)^2}+\sqrt{|\vec p|^2+(M-\Delta)^2}.
\end{eqnarray*}
Since the leading terms of the asymptotic expansion (\ref{16}) do not
depend on $\mu$, it is clear that the last integral in (\ref{013}) is
a convergent one. Other terms in (\ref{013}) form the unrenormalized
TDP (effective potential) at $\mu=0$ which is reduced after
renormalization procedure to the expression (\ref{25}). Hence, after
renormalization we obtain from (\ref{013}) the following finite
expression (evidently, it coincides with renormalized TDP
(\ref{19})):
 \begin{eqnarray}
\Omega^{ren} (M,\Delta)=V(M,\Delta)
-\int\frac{d^2p}{(2\pi)^2}\Big ({\cal E}_\Delta^++{\cal E}_\Delta^--
\sqrt{|\vec p|^2+(M+\Delta)^2}-\sqrt{|\vec p|^2+(M-\Delta)^2}\Big ), \label{24}
\end{eqnarray}
where $V(M,\Delta)$ is presented in (\ref{25}). The integral term in
(\ref{24}) can be explicitly calculated. As a result, we have
\begin{eqnarray}
12\pi\Omega^{ren}
(M,\Delta)&=&\frac{6M^2}{g_1}+\frac{6\Delta^2}{g_2}+
2\left (M+\sqrt{\mu^2+\Delta^2}\right )^3+2\left |M-\sqrt{\mu^2+\Delta^2}\right |^3\nonumber\\
&-&3t_+\left(M+\sqrt{\mu^2+\Delta^2}\right )+
3t_-\left |M-\sqrt{\mu^2+\Delta^2}\right |\nonumber\\
&-&\frac{3(\mu^2-M^2)\Delta^2}{\mu}\ln\left
|\frac{t_++\mu(M+\sqrt{\mu^2+\Delta^2})}{t_-+\mu
|M-\sqrt{\mu^2+\Delta^2}|}\right |, \label{23}
\end{eqnarray}
where $t_\pm=M\sqrt{\mu^2+\Delta^2}\pm\mu^2$. It is not so evident,
but at $\mu=0$ the expression (\ref{23}) for $\Omega^{ren}
(M,\Delta)$   coincides with $V(M,\Delta)$ (\ref{25}).

\section{Phase structure of the model at $T=0$}

As was mentioned above, the coordinates of the global minimum point
$(M_0,\Delta_0)$ of the TDP $\Omega^{ren}(M,\Delta)$ define the ground
state expectation values of auxiliary fields $\sigma (x)$ and
$\Delta (x)$. Namely, $M_0=\vev{\sigma(x)}$ and
$\Delta_0=\vev{\Delta(x)}$. The quantities $M_0$ and $\Delta_0$ are
usually called order parameters, or gaps, because they are
responsible for the phase structure of the model or, in other words,
for the properties of the model ground state (see also the comment
after (\ref{03})). Moreover, the gap $M_0$ is equal to the dynamical
mass of  one-fermionic excitations of the ground state. As a rule,
gaps depend on model parameters as well as on various external
factors. In our consideration the gaps $M_0$ and $\Delta_0$
are certain functions of the free model parameters $g_1$ and $g_2$ and
such external factors as chemical potential $\mu$ and temperature $T$.
\begin{figure}
%----figure 1,2
 \includegraphics[width=0.45\textwidth]{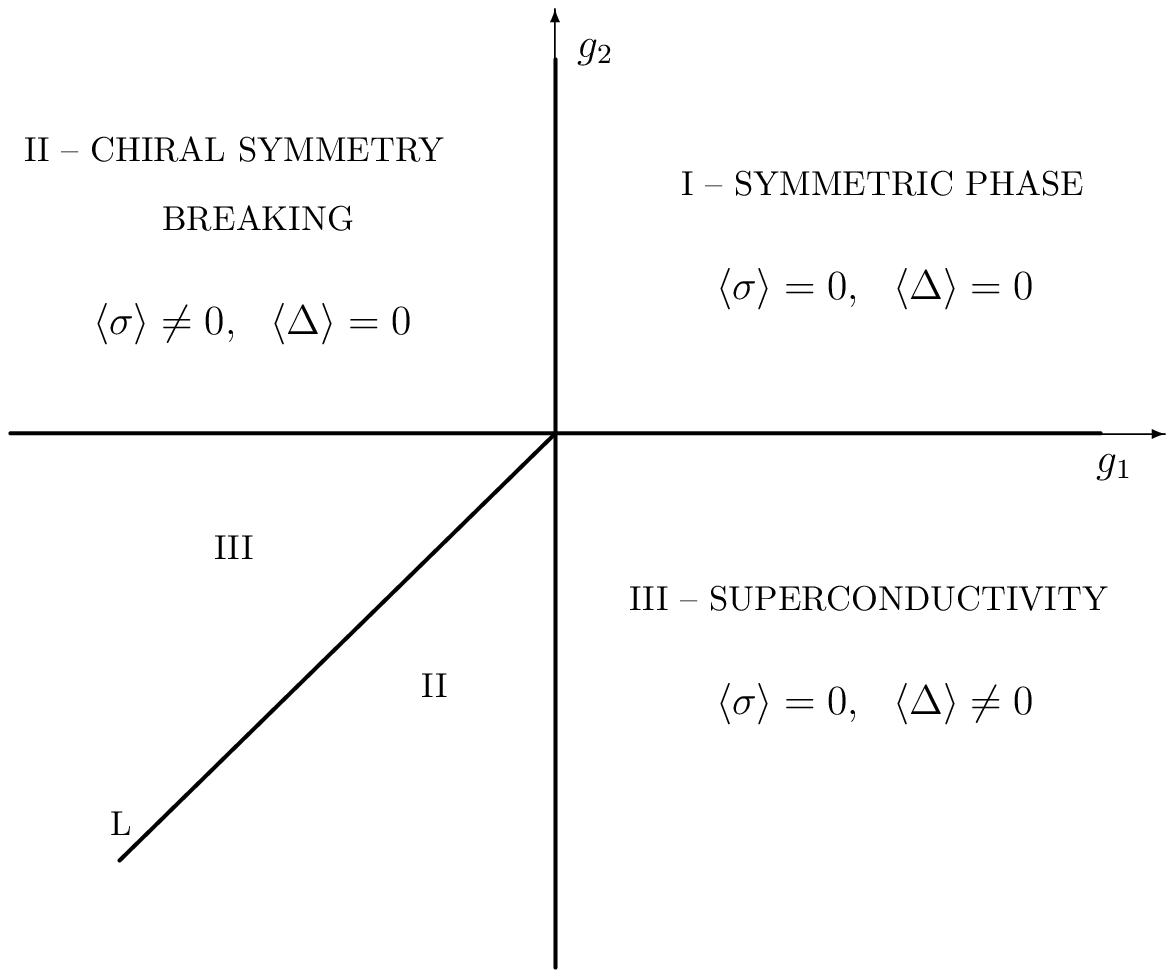}
 \hfill
 \includegraphics[width=0.45\textwidth]{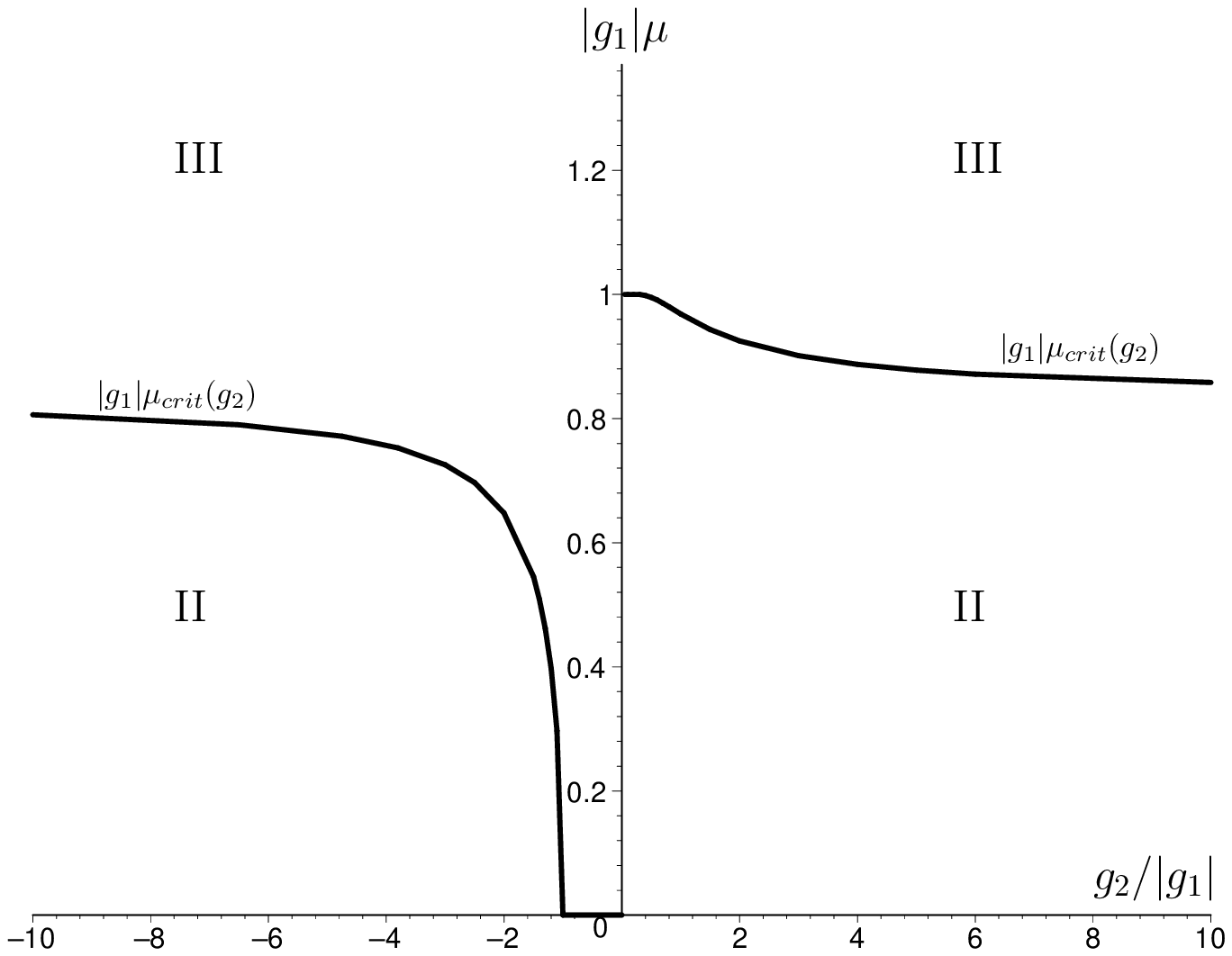}\\
\parbox[t]{0.45\textwidth}{
 \caption{The $(g_1,g_2)$-phase portrait of the model at $\mu=0$.
 The shorthands I, II and III denote the symmetric, the chiral symmetry
 breaking and the superconducting phases, respectively. In the phase II
 $\vev{\sigma}=-1/g_1$. In the phase III $\vev{\Delta}=-1/g_2$. On the line
 L$\equiv\{(g_1,g_2):g_1=g_2\}$, where $g_{1,2}<0$, the TDP minima corresponding to the phase II and III are equivalent.
%there is a
 %coexistence of the phases II and III.
}
 }\hfill
\parbox[t]{0.45\textwidth}{
\caption{The $(\mu,g_2)$-phase portrait of the model and critical
chemical potential $\mu_{crit}(g_2)$ vs $g_2$ at arbitrary fixed
$g_1<0$. At each point  $\mu=\mu_{crit}(g_2)\ne 0$ there is a first
order phase transition from the chiral symmetry breaking phase II to
the superconducting phase III.}   }
\end{figure}

\subsection{The case $\mu=0$}

First of all, let us discuss the phase structure of the model (1) in
the simplest case when $\mu=0$ and $T=0$. The corresponding TDP is
given in (\ref{25}) by the function $V(M,\Delta)$. Since the global
minimum of this function was already investigated in
\cite{Zhukovsky:2000yd}, although in the framework of another
(2+1)-dimensional GN model, we present at once the phase structure
of the initial model (1) at $\mu=0$ (see Fig. 1).

In Fig. 1 the phase portrait of the model is depicted depending on
the values of the free model parameters $g_1$ and $g_2$. There the
plane $(g_1,g_2)$ is  divided into several areas.  In each area one
of the phases I, II or III is implemented. In the phase I, i.e. at
$g_1>0$ and $g_2>0$, the global minimum of the effective potential
$V(M,\Delta)$ is arranged at the origin. So in this case we have
$M_0=\vev{\sigma(x)}=0$ and $\Delta_0=\vev{\Delta(x)}=0$. As a
result, in the phase I both discrete chiral and continuous
electromagnetic $U(1)$ symmetries remain intact and fermions are
massless. Due to this reason the phase I is called symmetric. In
the phase II, which is allowed only for $g_1<0$, at the global
minimum point $(M_0,\Delta_0)$ the relations $M_0=-1/g_1$ and
$\Delta_0=0$ are valid. So in this phase chiral symmetry is
spontaneously broken down and fermions acquire dynamically the mass
$M_0$. Finally, in the superconducting phase III, where $g_2<0$, we
have the following values for the gaps $M_0=0$ and
$\Delta_0=-1/g_2$.

Note also that if $g_1=g_2\equiv g$ and, in addition, $g<0$ (it is
just the line L in Fig. 1), then the effective potential (\ref{25})
has two equivalent global minima. The first one, the point
$(M_0=-1/g,\Delta_0=0)$, corresponds to a phase with chiral symmetry
breaking. The second one, i.e. the point $(M_0=0,\Delta_0=-1/g)$,
corresponds to superconductivity. 

Clearly, if the cutoff parameter $\Lambda$ is fixed, then the phase
structure of the model can be described in terms of bare coupling
constants $G_1, G_2$ instead of finite quantities $g_1, g_2$.
Indeed, let us first introduce a critical value of the couplings,
$G_c=\frac{\pi^2}{8\Lambda\ln(1+\sqrt{2})}$. Then, as it follows
from Fig. 1 and (\ref{18}), at $G_1<G_c$ and $G_2<G_c$ the symmetric
phase I of the model is located. If $G_1>G_c$, $G_2<G_c$ ($G_1<G_c$,
$G_2>G_c$), then the chiral symmetry broken phase II (the
superconducting phase III) is realized. Finally, let us suppose that
both $G_1>G_c$ and $G_2>G_c$. In this case at $G_1>G_2$ ($G_1<G_2$)
we have again the chiral symmetry broken phase II (the
superconducting phase III).
\begin{figure}
%----figure 3,4
 \includegraphics[width=0.45\textwidth]{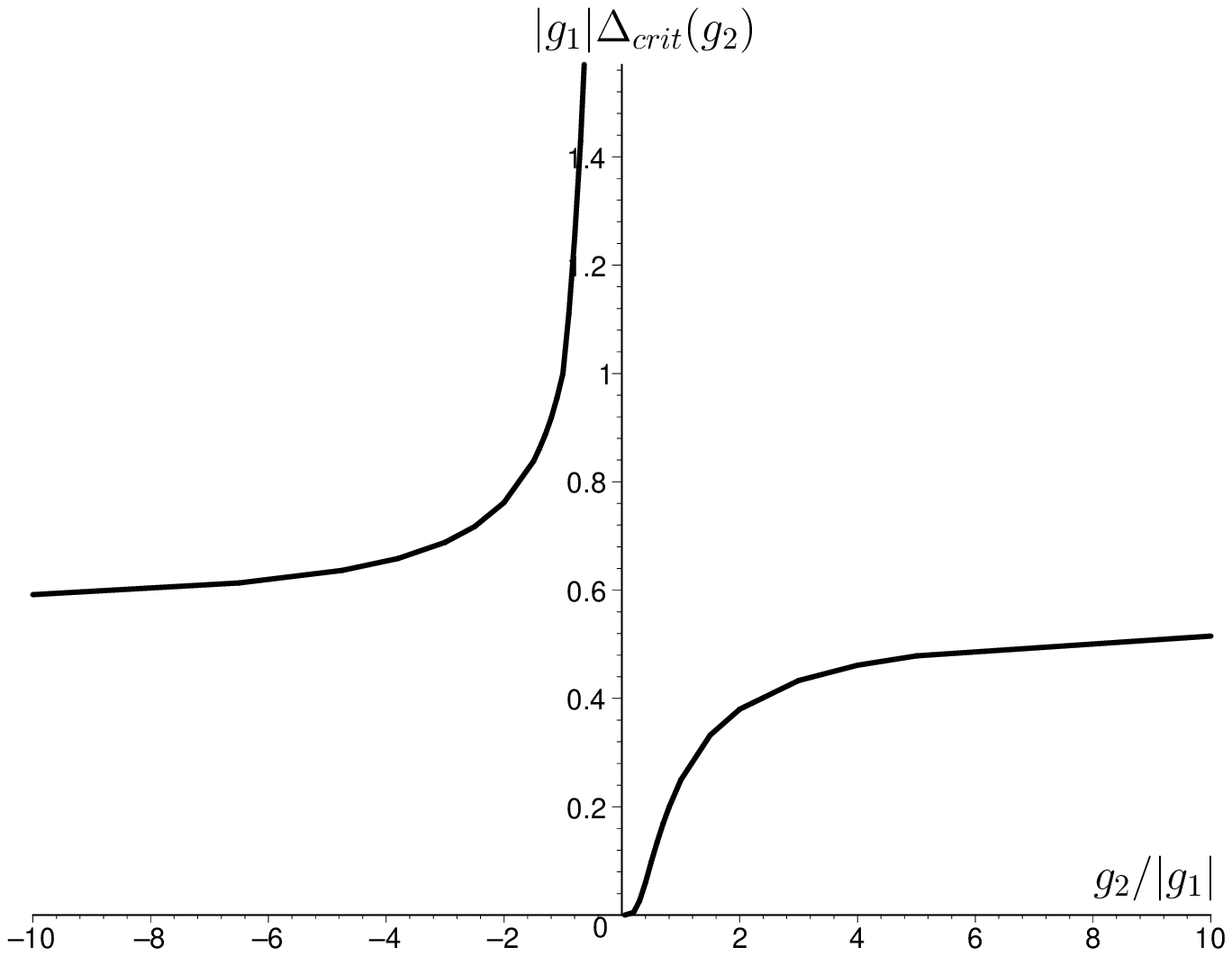}
 \hfill
 \includegraphics[width=0.45\textwidth]{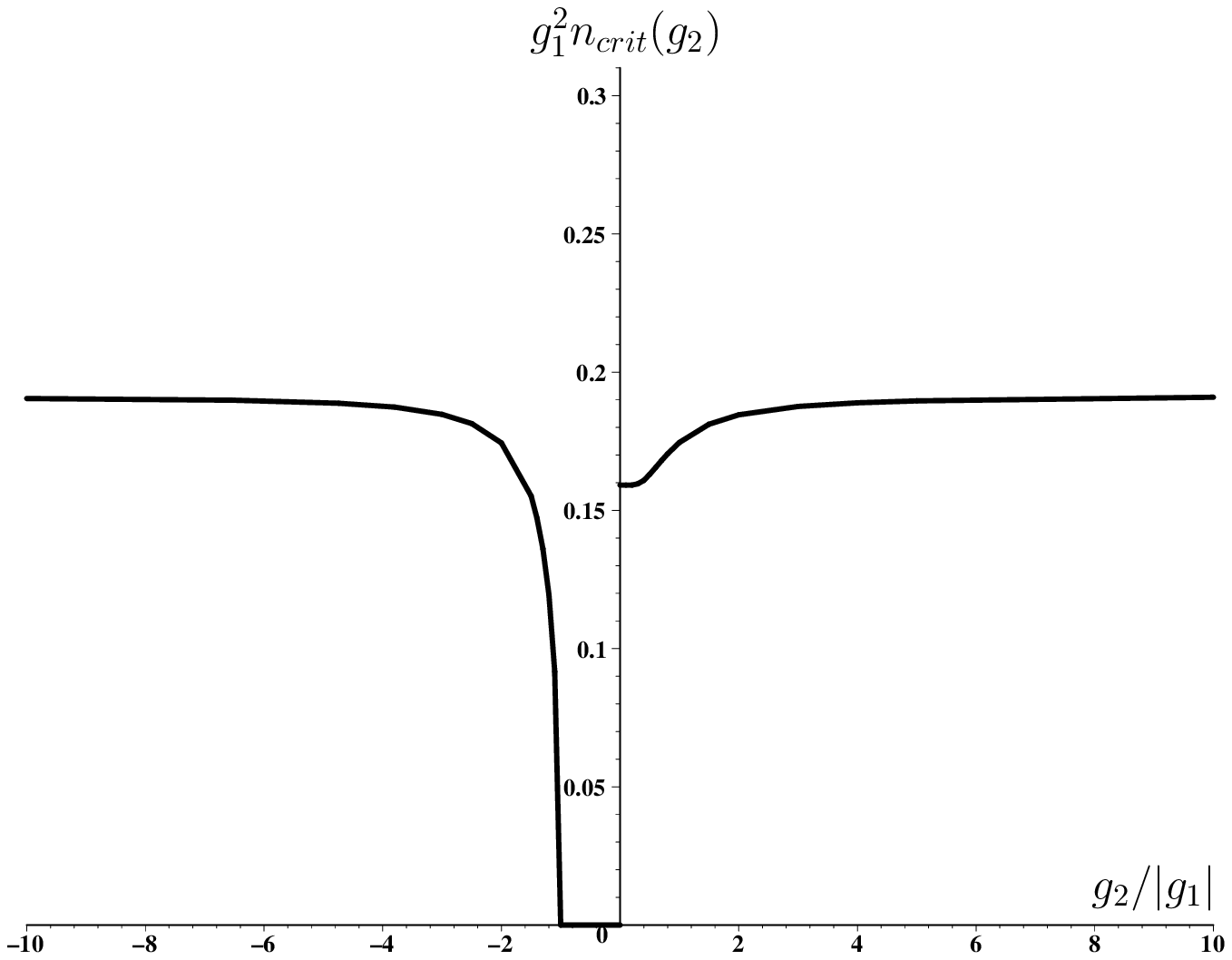}\\
\parbox[t]{0.45\textwidth}{
 \caption{Superconducting gap $\Delta_0=\Delta_{crit}(g_2)$ vs $g_2$ which is generated at the critical point, i.e. at $\mu=\mu_{crit}(g_2)$, at arbitrary fixed $g_1<0$.}
 }\hfill
\parbox[t]{0.45\textwidth}{
\caption{Particle density $n=n_{crit}(g_2)$ vs $g_2$ which is generated
 at the critical point, i.e. at $\mu=\mu_{crit}(g_2)$, at arbitrary fixed $g_1<0$.
 At $\mu<\mu_{crit}(g_2)$ the particle density $n$ is equal to zero. }    }
\end{figure}
\begin{figure}
%----figure 5,6
\includegraphics[width=0.45\textwidth]{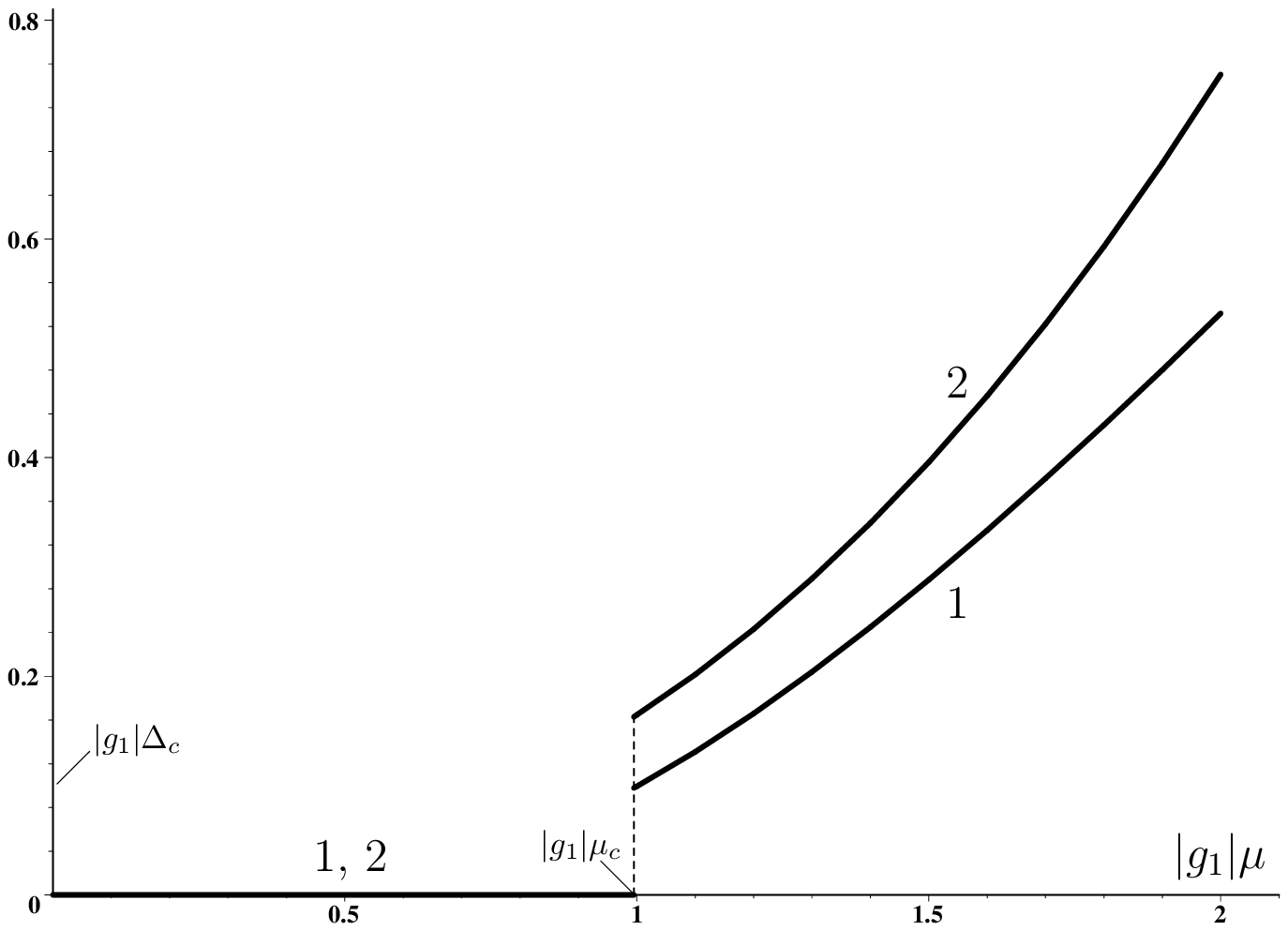}
\hfill
\includegraphics[width=0.45\textwidth]{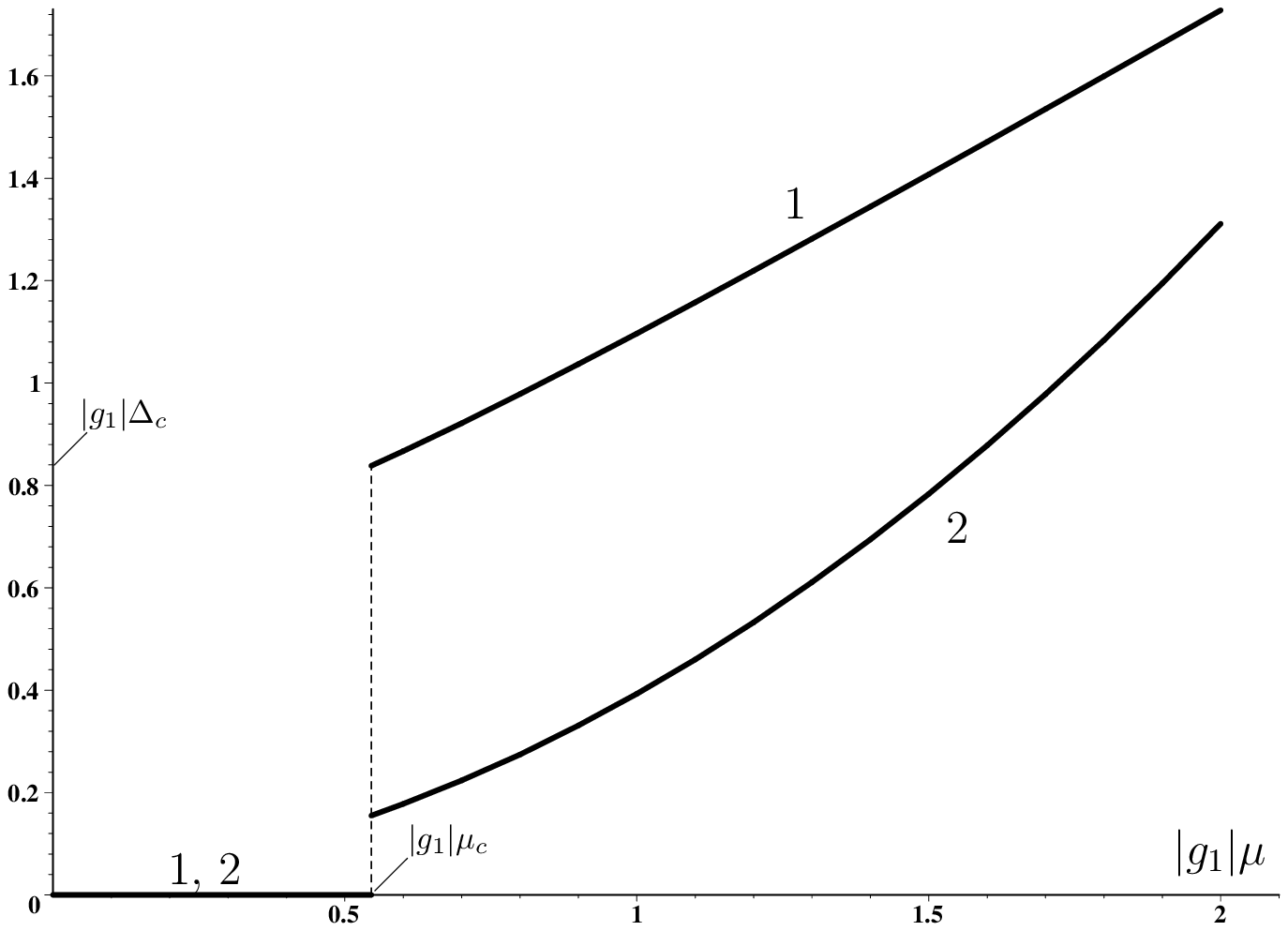}\\
\parbox[t]{0.45\textwidth}{
\caption{Superconducting gap $\Delta_0$ and particle  density $n$ vs
$\mu$ at arbitrary fixed $g_1<0$ and $g_2=0.5|g_1|$. Curves 1 and 2 are the
plots of the dimensionless quantities $|g_1|\Delta_0$ and
$|g_1|^2n$, correspondingly. Here
$|g_1|\mu_c=|g_1|\mu_{crit}(g_2=0.5|g_1|)\approx0.995$ and
$|g_1|\Delta_c=|g_1|\Delta_{crit}(g_2=0.5|g_1|)\approx0.098$. }
 }\hfill
\parbox[t]{0.45\textwidth}{
\caption{Superconducting gap $\Delta_0$ and particle density $n$ vs
$\mu$ at arbitrary fixed $g_1<0$ and $g_2=-1.5|g_1|$. Curves 1 and 2 are the plots of the dimensionless quantities $|g_1|\Delta_0$ and
$|g_1|^2n$, respectively. Here
$|g_1|\mu_c=|g_1|\mu_{crit}(g_2=-1.5|g_1|)\approx 0.545$ and
$|g_1|\Delta_c=|g_1|\Delta_{crit}(g_2=-1.5|g_1|)\approx 0.838$. }  }
\end{figure}
\label{mu0}

\subsection{Consideration of the chemical potential}
\label{munot0}

In this section we study the influence of the chemical potential
$\mu> 0$ on the phase structure of the model (1) (temperature is
still vanishing). Numerical and analytical investigations of the TDP
(\ref{23}) show that its minimum points are of the form $(M\not
=0,\Delta=0)$, $(M=0,\Delta\not=0)$ or $(M=0,\Delta=0)$ only. So to
study the properties of the global minimum point of the function
(\ref{23}) it is enough to consider its reductions on the $M$- and
$\Delta$-axes, where the TDP (\ref{23}) becomes
\begin{eqnarray}
12\pi\Omega^{ren} (M,\Delta)\Big |_{\Delta=0}\equiv
12\pi\omega_1(M)&=&\frac{6M^2}{g_1}+
2\left (M+\mu\right )^3+2\left |M-\mu\right |^3\nonumber\\
&-&3\mu\left(M+\mu\right )^2+
3\mu (M-\mu)\left |M-\mu\right |,\label{26}\\
12\pi\Omega^{ren} (M,\Delta)\Big |_{M=0}\equiv
12\pi\omega_2(\Delta)&=&\frac{6\Delta^2}{g_2}+
4(\mu^2+\Delta^2)^{3/2}-6\mu^2\sqrt{\mu^2+\Delta^2}\nonumber\\&-&
3\mu\Delta^2\ln\left
(\frac{(\mu+\sqrt{\mu^2+\Delta^2})^2}{\Delta^2}\right
),\label{27}
\end{eqnarray}
respectively. Comparing the minima of the functions (\ref{26}) and (\ref{27}), it is possible to find the global minimum point of the whole TDP (\ref{23}) and its dependence on the model parameters $\mu$, $g_1$
and $g_2$, i.e. to determine the phase structure of the model. In
addition, in the present section we will study the behavior of a
particle density $n$ in different phases when $\mu$ varies,
\begin{eqnarray}
n=-\frac{\partial\Omega^{ren} (M,\Delta)}{\partial\mu}\Big
|_{M=M_0,\Delta=\Delta_0}.\label{28}
\end{eqnarray}
Since the global minimum point (GMP) of the TDP (\ref{23}) coincides
with the GMP either of the function $\omega_1(M)$ (\ref{26}) or the
function $\omega_2(\Delta)$ (\ref{27}), it is clear that in the
chirally broken phase $\Delta_0=0$ and the gap $M_0$ does not depend
on the parameter $g_2$. Correspondingly, in the superconducting
phase we have $M_0=0$ and the gap $\Delta_0$ does not depend on the
parameter $g_1$. So, one can use the following expressions for the
particle density in the chiral symmetry broken II and
superconducting III phases:
\begin{eqnarray}
n\big |_{\rm phase~
II}&=&-\frac{\partial\omega_1(M)}{\partial\mu}\Big
|_{M=M_0}=\frac{1}{2\pi}(\mu^2-M_0^2)\theta (\mu-M_0),\label{29}\\
n\big |_{\rm phase~
III}&=&-\frac{\partial\omega_2(\Delta)}{\partial\mu}\Big
|_{\Delta=\Delta_0}= \frac{1}{2\pi}\left
[\mu\sqrt{\mu^2+\Delta^2_0}+\Delta^2_0\ln\frac{\mu+\sqrt{\mu^2+\Delta^2_0}}{\Delta_0}\right
],\label{30}
\end{eqnarray}
where $\theta(x)$ is the Heaviside step-function.

{\bf The case $g_1<0$.} First of all, let us suppose that $g_1$ is fixed and negative, i.e.
$g_1<0$. Then it is easy to show that for arbitrary value of $g_2$
there exists a critical chemical potential $\mu_{crit}(g_2)$ (see
Fig. 2) \footnote{All the Figs 2-10 are drawn in terms of dimensionless
quantities which are obtained after multiplication of appropriate
powers of $|g_1|$ with corresponding dimensional quantities. For
example, there instead of $\mu$, $\Delta_0$, $g_2$ we use their
dimensionless analogies $|g_1|\mu$, $|g_1|\Delta_0$, $g_2/|g_1|$.
Instead of particle density $n$ the dimensionless quantity $g_1^2n$
is depicted there etc. }
 such that at $\mu<\mu_{crit}(g_2)$ the system is in the chiral symmetry breaking
phase II (if $\mu_{crit}(g_2)>0$), and it is in the superconducting phase III at
$\mu>\mu_{crit}(g_2)$. In other words, if $\mu<\mu_{crit}(g_2)\ne 0$,
then the global minimum of the TDP (\ref{23}) lies at the point
$(M_0=-1/g_1,\Delta_0=0)$ which does not depend on $\mu$ in the
interval $0<\mu<\mu_{crit}(g_2)$. However, at $\mu=\mu_{crit}(g_2)$
it jumps to the point $(M_0=0,\Delta_0=\Delta_{crit}(g_2))$, where
$\Delta_{crit}(g_2)$ vs $g_2$ is depicted in Fig. 3. Hence, at the
critical point $\mu=\mu_{crit}(g_2)$ a first order phase transition
occurs and a superconducting gap $\Delta_0=\Delta_{crit}(g_2)$ is
dynamically generated. It turns out that $\Delta_0$ vs $\mu$ is an increasing function in the interval $\mu>\mu_{crit}(g_2)$. In particular, the behavior $\Delta_0$ vs $\mu$ is presented in Fig. 5 (at $g_2=0.5|g_1|$), Fig. 6 (at $g_2=-1.5|g_1|$) and Fig. 7 (at $g_2=-0.5|g_1|$) as the curve 1.

Moreover, it is clear from Fig. 2 that at $\mu<\mu_{crit}(g_2)$,
i.e. in the phase II, the particle density is equal to zero. To
explain this circumstance, recall that in the phase II the gap $M_0$
is equal to $1/|g_1|$. So, for all $g_2$-values the relation
$\mu_{crit}(g_2)<M_0$ is valid (see Fig. 2). As a result, throughout
the phase II, where $\mu<\mu_{crit}(g_2)$, we have $\mu<M_0$ and
hence, as it follows from the relation (\ref{29}), the zero particle
density, $n=0$. However, when $\mu$ reaches its critical value,
$\mu=\mu_{crit}(g_2)$, the nonzero particle density $n_{crit}(g_2)$
is generated dynamically in the system (see Fig. 4). Further growth
of the chemical potential is accompanied by increase of the particle
density $n$ vs $\mu$. (Evidently, in this case the particle density
must be calculated with the help of the expression (\ref{30}).) For
example, in Figs 5--7 at the same
representative relations between $g_1$ and $g_2$ the particle
density $n$ vs $\mu$ is depicted as a monotonically increasing curve
2.

Finally recall that at $\mu=0$ the two phases, II and III, have equivalent minima of the TDP only  at negative values of $g_1=g_2$ (it is the line L in Fig. 1). It turns out that for arbitrary fixed $g_1<0$ and at growing
chemical potential, this property of the TDP is also allowed but
in a much more extensive $g_2$-region. Indeed, as our analysis shows
in this case, if $g_2>0$ or $g_2<g_1$ then at $\mu=\mu_{crit}(g_2)$
(see Fig. 2) the TDP has two equivalent minima, corresponding to
these phases. As a result, for these values of $g_1$ and $g_2$ there
is a coexistence of chirally broken and superconducting phases at
$\mu=\mu_{crit}(g_2)$. In this case, when viewed from the side we
have the following picture of phase transitions in the system. At
rather small values of $\mu$ the ground state of the system is an
empty space (particle density is zero). If fermions are created in
this state, they have a mass equal to $M_0=-1/g_1$, i.e. the ground
state corresponds to a chirally broken phase II. Then, if $\mu$
reaches the critical value $\mu=\mu_{crit}(g_2)$, bubbles of a new
phase III appear in the empty space. Inside each bubble the particle
density $n$ is nonzero and equal to $n_{crit}(g_2)$ (see Fig. 4).
\begin{figure}
%----figure 7,8
 \includegraphics[width=0.45\textwidth]{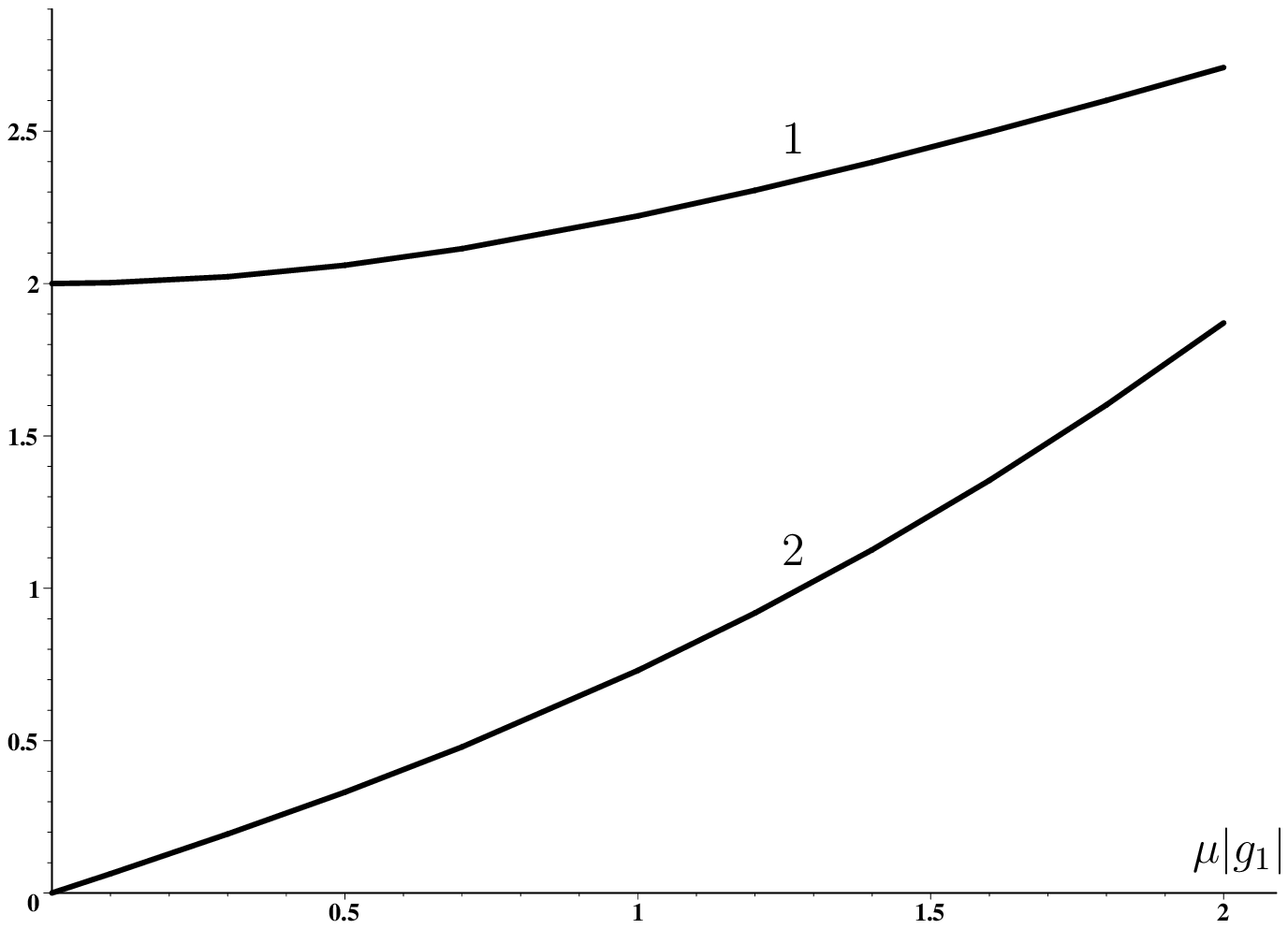}
 \hfill
 \includegraphics[width=0.45\textwidth]{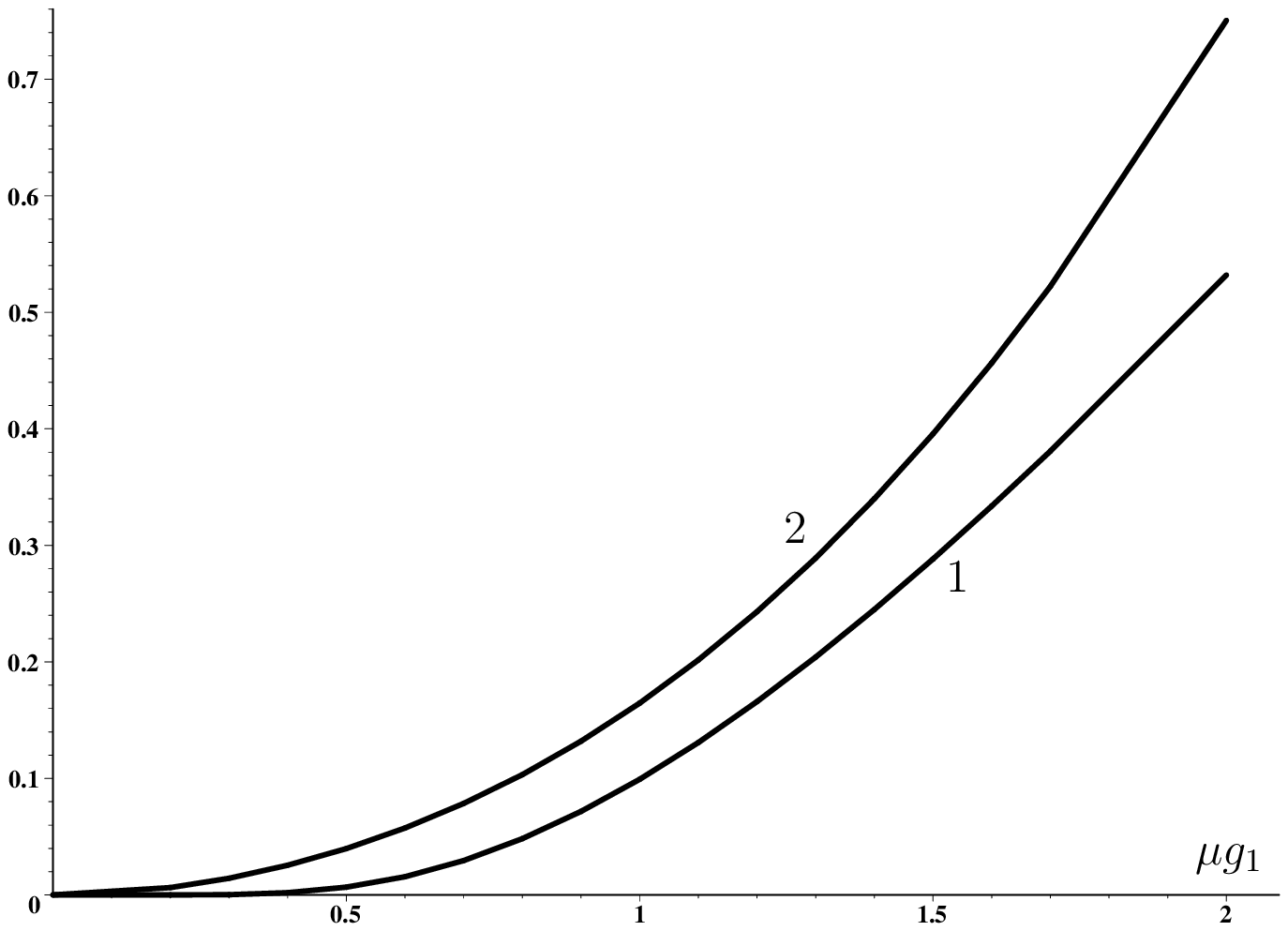}\\
\parbox[t]{0.45\textwidth}{
 \caption{Superconducting gap $\Delta_0$ and particle density $n$ vs
$\mu$ at arbitrary fixed $g_1$ (both at $g_1<0$ and $g_1>0$) as well as at
$g_2=-0.5|g_1|$. Curves 1 and 2 are the plots of the dimensionless
quantities $|g_1|\Delta_0$ and $6|g_1|^2n$, respectively.  }
 }\hfill
\parbox[t]{0.45\textwidth}{
\caption{Superconducting gap $\Delta_0$ and particle density $n$ vs
$\mu$ at arbitrary fixed $g_1>0$ as well as at $g_2=0.5g_1$. Curves 1 and 2
are the plots of the dimensionless quantities $|g_1|\Delta_0$ and
$|g_1|^2n$, respectively. }  }
\end{figure}

{\bf The case $g_1>0$.} Now the model phase structure consideration
for a  positive $g_1$-values is in order.  Recall, in this case we
have a rather weak attractive interaction in the chiral channel,
i.e. $G_1<G_c$. Evidently, if in addition $g_2<0$, then in this case
the superconducting phase is realized for
arbitrary values of $\mu\ge 0$. The behavior of the gap $\Delta_0$
and particle density $n$ vs $\mu$ in this branch of the
superconducting phase is given in Fig. 7 in the particular case
$g_2=-0.5g_1$ for $g_1>0$. Moreover, as it is clear from Fig. 7, the
same behavior for $\Delta_0$ and $n$ vs $\mu$ remains valid for the
case $g_2=-0.5|g_1|$ and negative values of $g_1$. To explain this
fact, it is necessary to take into account the remark made after
formula (\ref{28}) that the superconducting gap does not depend on
the coupling $g_1$ but only on $g_2$ one. So, it is no wonder that
the plots of $\Delta_0$ and $n$ are not changed when the parameter
$g_1$ changes the sign.

Recall, if both $g_1>0$ and $g_2>0$, then we have at $\mu=0$ the
phase I without any symmetry breaking, where the gaps $\Delta_0$ and
$M_0$ vanishes (see Fig. 1). However, our analysis shows that at
arbitrary small nonzero $\mu$ the global minimum point of the TDP
(\ref{23}) moves from the point $(M_0=0,\Delta_0=0)$ to the
following one $(M_0=0,\Delta_0\ne 0)$. Hence, at positive values of
$g_1$ and $g_2$ a continuous second order phase transition occurs
from symmetric phase I to superconducting one III when chemical
potential acquires an arbitrary small nonzero value. The typical
behavior of the gap $\Delta_0$ and particle density $n$ vs $\mu$ in
this superconductivity region is depicted in Fig. 8. Comparing Figs.
7 and 8, we see that at the same value of $\mu$ the gap $\Delta_0$
and particle density $n$ are much greater in the case $g_1>0$,
$g_2<0$, than in the case $g_1>0$, $g_2>0$. To support this
statement we draw in Figs. 9 and 10 the plots of the gap $\Delta_0$
and particle density $n$ vs $g_2$ in two different regions $g_2<0$
and $g_2>0$, respectively, at the particular value of the
chemical potential, $\mu=0.5/g_1$.

We see that at $g_1>0$, i.e. at $G_1<G_c$, the chiral symmetry
breaking is absent but the Cooper  pairing phase occurs at any
$\mu>0$. To explain this different behavior, one can use the
following very naive physical arguments. Since at $\mu>0$ we have a
nonzero particle density (see, e.g., in Fig. 8), there is a Fermi sea
of particles with energies less or equal to $\mu$ (Fermi surface).
Evidently, in this case there is no energy cost for creating a pair
of particles with opposite momenta just over the Fermi surface.
Then, due to an arbitrary weak attraction between these particles
($G_2>0$), the Cooper pair is formed and $U(1)$ symmetry is
spontaneously broken, as a result of Bose--Einstein condensation of
Cooper pairs. Note, since in the energy spectrum of fermions the gap
$\Delta\ne 0$ appears (see in Fig. 8), rather small external forces
are not able to destroy the superconducting condensate and it is a
stable one.

    Concerning the chiral symmetry breaking in this case, it is clear
that a particle and a hole with opposite momenta can also be created
without any energy cost in the system. Moreover, there is also an
attraction between a particle and a hole. However, since the nonzero
gap $M$ does not appear in the energy spectrum at sufficiently small
$G_1<G_c$, the particle--hole pairing in this case is rather a
weakly bounded resonance, which, unlike a stable pair, could be easily
destroyed  by an  arbitrary small external influence. So, no
stable Bose--Einstein condensate of these pairs does appear and
chiral symmetry remains intact. (For a more detailed discussion on
possible types of pairing in dense (quark) fermionic matter see,
e.g., in \cite{kojo}.)

In summary, we can say that at $T=0$ chemical potential
induces superconductivity in the model for arbitrary relations between coupling constants $g_{1,2}$ (or, equivalently, $G_{1,2}$).
\begin{figure}
%----figure 9,10
 \includegraphics[width=0.45\textwidth]{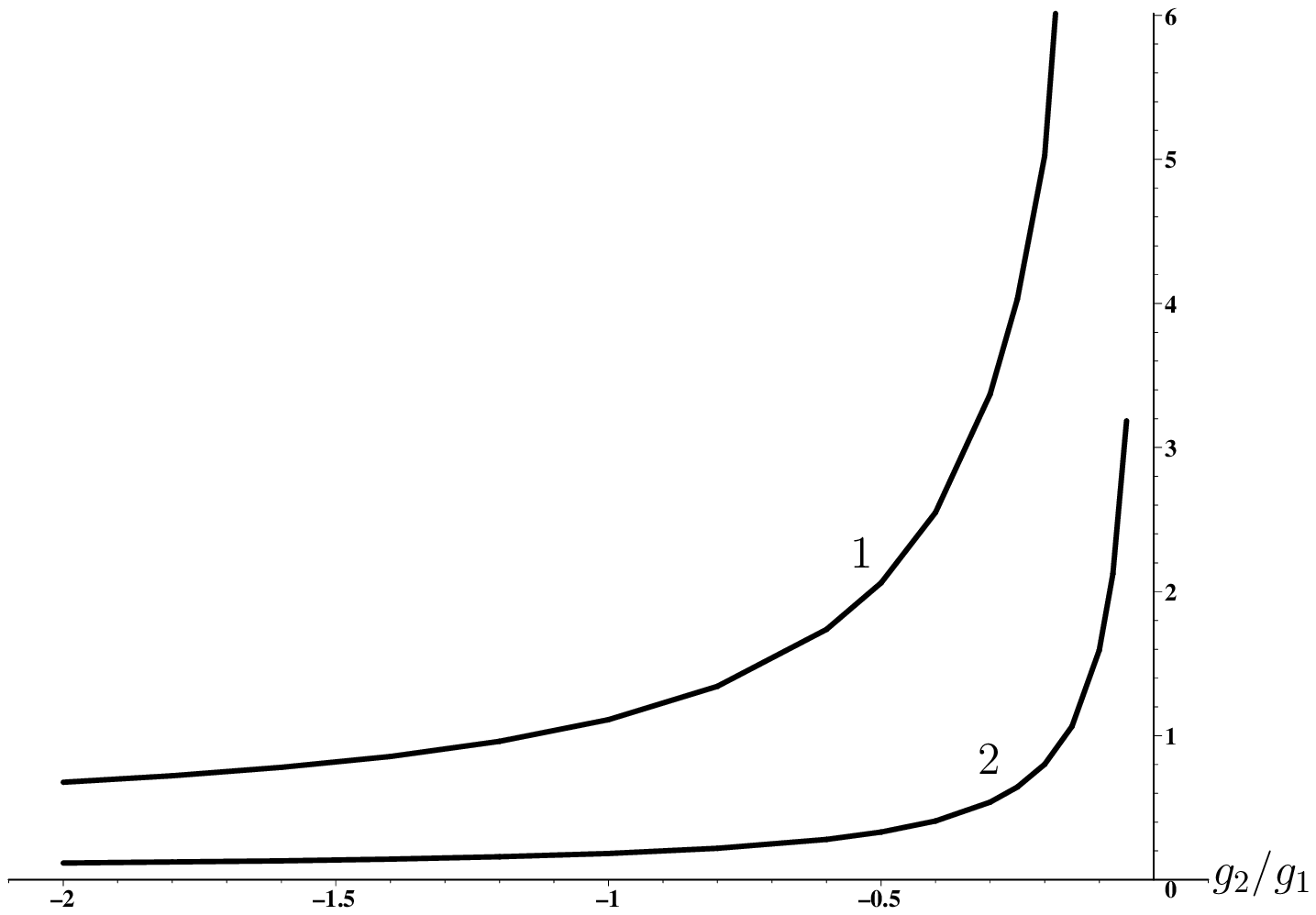}
 \hfill
 \includegraphics[width=0.45\textwidth]{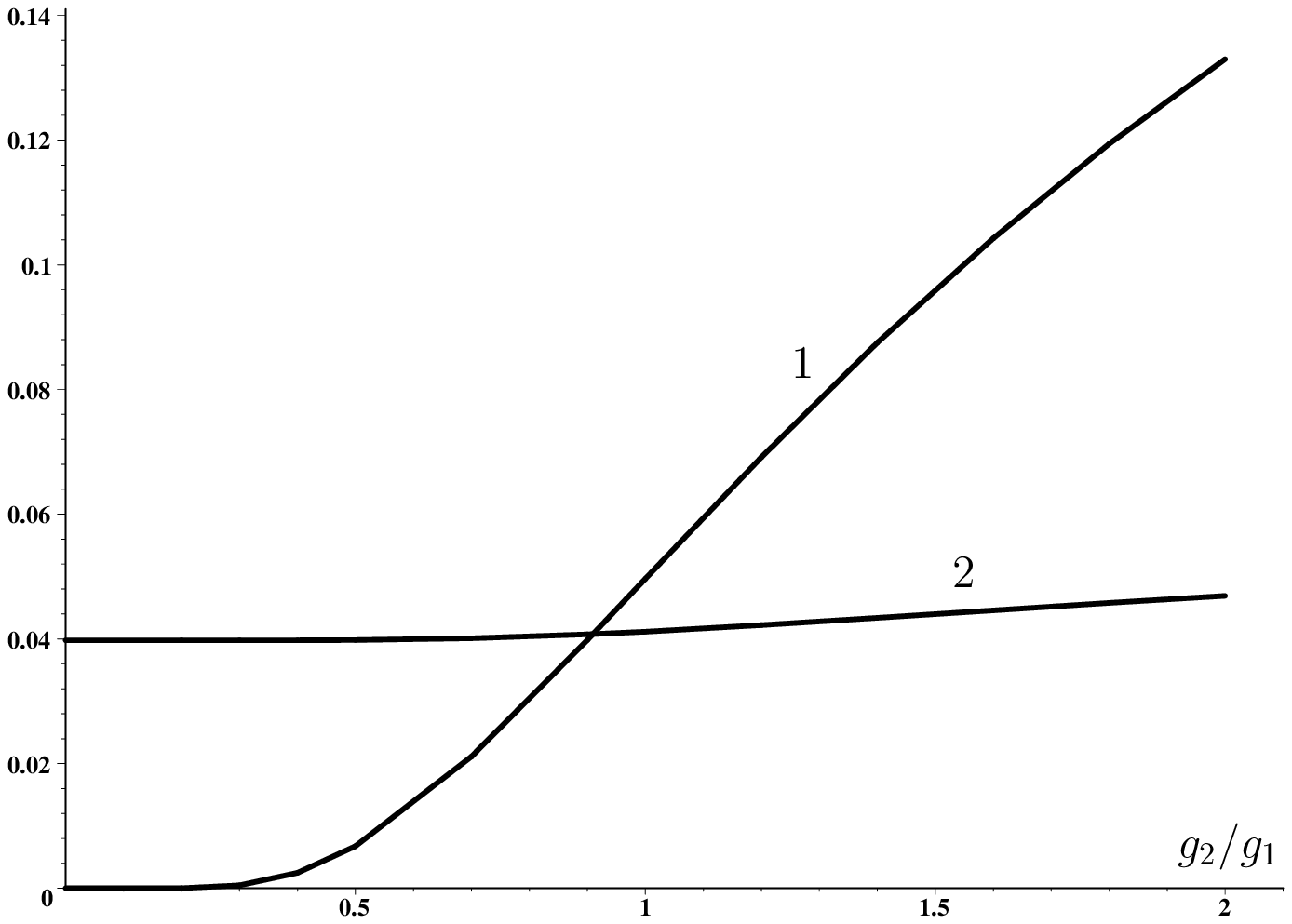}\\
\parbox[t]{0.45\textwidth}{
 \caption{Superconducting gap $\Delta_0$ and particle density $n$ vs
$g_2<0$ at arbitrary fixed $g_1>0$ and $\mu=0.5/g_1$. Curves 1 and 2
are the plots of the dimensionless quantities $g_1\Delta_0$ and
$g_1^2n$, respectively. }
 }\hfill
\parbox[t]{0.45\textwidth}{
\caption{Superconducting gap $\Delta_0$ and particle density $n$ vs
$g_2>0$ at arbitrary fixed $g_1>0$ and $\mu=0.5/g_1$. Curves 1 and 2
are the plots of the dimensionless quantities $g_1\Delta_0$ and
$g_1^2n$, respectively. }  }
\end{figure}

\section{Finite temperature}
\label{t}

 Now let us study the influence of both temperature $T$ and
chemical potential $\mu$ on the phase structure of the model. It is
well known (see,  e.g., in \cite{barducci}) that in $d$ space
dimensions (in our case, evidently, $d=2$) the transition probability
from one degenerated minimum of the TDP to another is proportional to $\exp
(-N\beta L^{d-2})$, where $L$ is the linear size of the system and
$\beta$ is the inverse temperature, $\beta=1/T$. It follows from
this expression that at $d=2$ the transition probability
is zero even at finite $N$ if $T=0$. This leads to the fact that a continuous symmetry can be spontaneously broken in any planar systems at $T=0$. (Hence, our consideration of superconducting phase transitions
performed at $T=0$ in the previous section is valid for arbitrary
values of $N$.) However, if $T\ne 0$, then transition probability in the above expression does not vanish at finite $N$. This circumstance
ensures vanishing of the order parameter and, as a result, might
lead to a prohibition for spontaneous symmetry breaking in $d=2$
spatial dimensions at finite $N$ and $T\ne 0$. However, if $N\to\infty$ the transition probability vanishes and the spontaneous symmetry breaking is
allowed. Just this assumption, i.e the same as in
\cite{chodos,abreu}, is used in the following consideration, where
we study the temperature dependent superconducting phase transitions
in the leading order of large-$N$ expansion technique.
\begin{figure}
%----figure 11,12
 \includegraphics[width=0.45\textwidth]{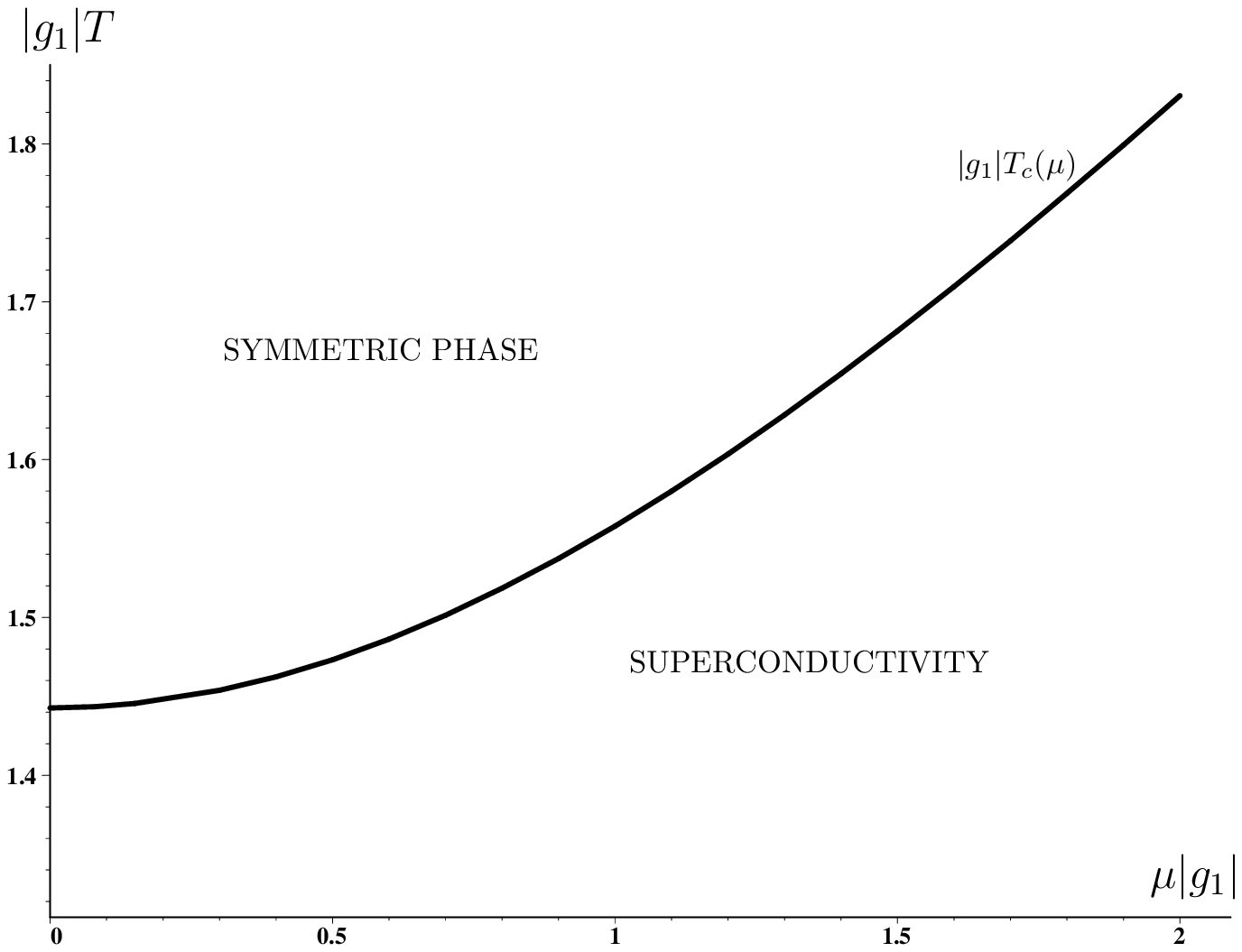}
 \hfill
 \includegraphics[width=0.45\textwidth]{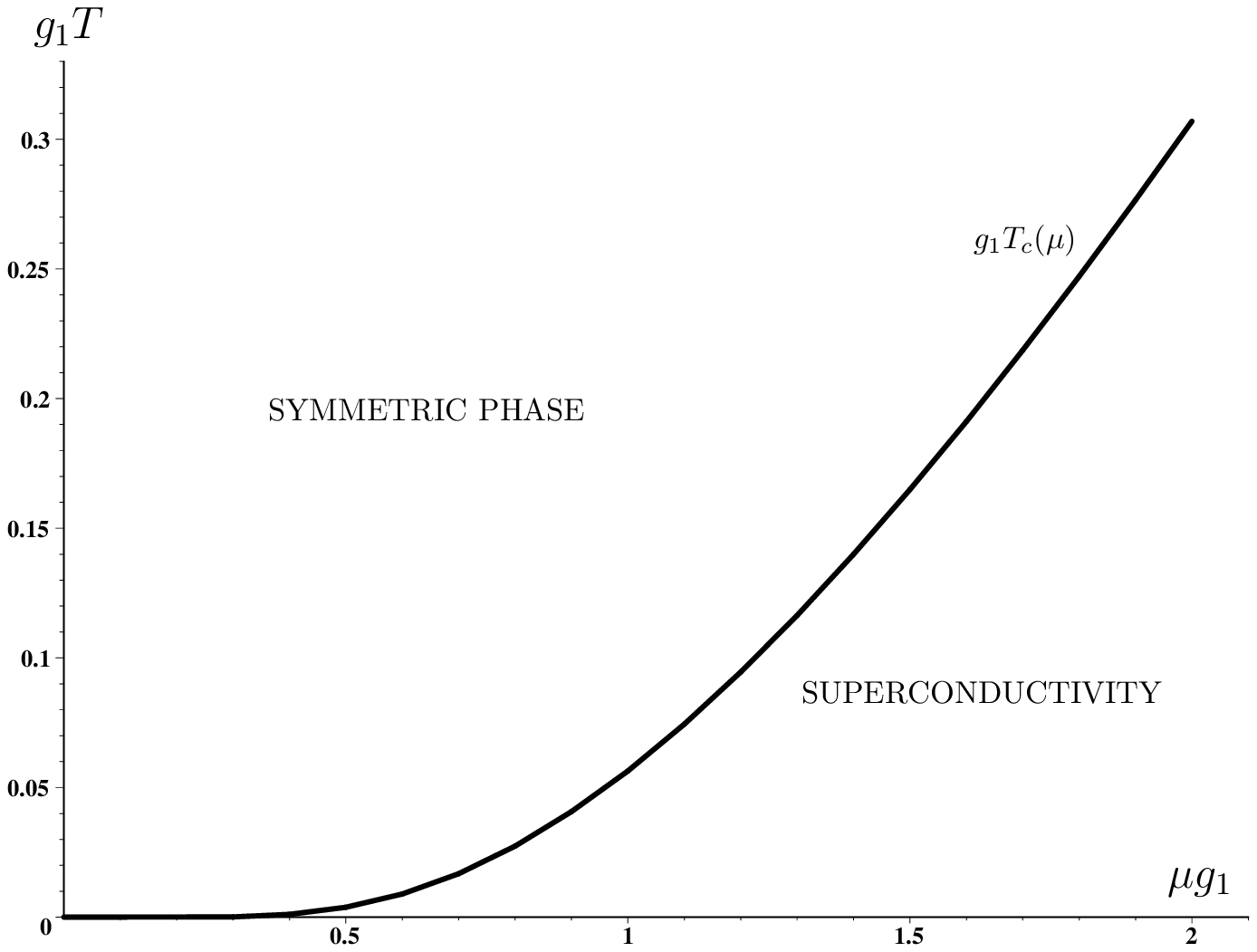}\\
\parbox[t]{0.45\textwidth}{
\caption{$(\mu,T)$-phase diagram of the model at $g_2=-0.5|g_1|$ and
arbitrary fixed $g_1$ both at $g_1<0$ and $g_1>0$. }
 }\hfill
\parbox[t]{0.45\textwidth}{
\caption{$(\mu,T)$-phase diagram of the model at arbitrary fixed
$g_1>0$  and at $g_2=0.5g_1$. }  }
\end{figure}

In this case, in order
to get the corresponding (unrenormalized) thermodynamic potential
$\Omega_{\scriptscriptstyle{T}}(M,\Delta)$ one can simply start from
the expression for the TDP at zero temperature (\ref{12}) and
perform the following standard replacements:
\begin{eqnarray}
\int_{-\infty}^{\infty}\frac{dp_0}{2\pi}\big (\cdots\big )\to
iT\sum_{n=-\infty}^{\infty}\big (\cdots\big ),~~~~p_{0}\to
p_{0n}\equiv i\omega_n \equiv i\pi T(2n+1),~~~n=0,\pm 1, \pm 2,...,
\label{190}
\end{eqnarray}
i.e. the $p_0$-integration should be replaced by the summation over
Matsubara frequencies $\omega_n$. Summing over
Matsubara frequencies in the obtained expression  (the corresponding
technique is presented, e.g., in \cite{jacobs}), one can find for
the TDP
\begin{eqnarray}
\Omega_{\scriptscriptstyle{T}}(M,\Delta)\!
&=&\frac{M^2}{4G_1}+\frac{\Delta^2}{4G_2}-\int_{-\infty}^{\infty}\frac{d^2p}{(2\pi)^2}
\left ({\cal E}^+_{\Delta}+{\cal E}^-_{\Delta}\right
)-2T\int_{-\infty}^{\infty}\frac{d^2p}{(2\pi)^2}\ln\left (\big
[1+e^{-\beta {\cal E}^+_{\Delta}}\big ]\big [1+e^{-\beta {\cal
E}^-_{\Delta}}\big ]\right ), \label{1202}
\end{eqnarray}
where $\beta=1/T$ and ${\cal E}^\pm_{\Delta}$ are given in
(\ref{12}). Clearly, only the first integral in this expression
(which is the same as in the zero temperature case) is responsible
for ultraviolet divergency of the whole TDP (\ref{1202}). So,
regularizing the TDP (\ref{1202}) in the way as it was done in
(\ref{15}) for zero temperature TDP and then replacing $G_{1,2}\to
G_{1,2}(\Lambda)$ (see formula (\ref{18})), we can obtain in the
limit $\Lambda\to\infty$ a finite expression denoted as
$\Omega^{ren}_{\scriptscriptstyle{T}}(M,\Delta)$,
\begin{eqnarray}
\Omega^{ren}_{\scriptscriptstyle{T}}(M,\Delta)\!
&=&\Omega^{ren}(M,\Delta)-2T\int_{-\infty}^{\infty}\frac{d^2p}{(2\pi)^2}\ln\left
(\big [1+e^{-\beta {\cal E}^+_{\Delta}}\big ]\big [1+e^{-\beta {\cal
E}^-_{\Delta}}\big ]\right ), \label{123}
\end{eqnarray}
where $\Omega^{ren}(M,\Delta)$ is the zero temperature TDP
(\ref{23}). Numerical investigations show that all possible local
minima of the TDP $\Omega^{ren}_{\scriptscriptstyle{T}}(M,\Delta)$
are located in the lines $M=0$ or $\Delta=0$. So it is sufficient to
deal with corresponding restrictions of the TDP on these lines, i.e.
with the following functions,
\begin{eqnarray}
F_1(M)\equiv\Omega^{ren}_{\scriptscriptstyle{T}}(M,\Delta)\Big
|_{\Delta=0}\!
&=&\omega_1(M)-2T\int_{-\infty}^{\infty}\frac{d^2p}{(2\pi)^2}\ln\left
(\big [1+e^{-\beta (E+\mu)}\big ]\big [1+e^{-\beta |E-\mu|}\big ]\right )\nonumber\\
&=&\frac{M^2}{2\pi
g_1}+\frac{M^3}{3\pi}-2T\int_{-\infty}^{\infty}\frac{d^2p}{(2\pi)^2}
\ln\left
(\big [1+e^{-\beta (E+\mu)}\big ]\big [1+e^{-\beta (E-\mu)}\big ]\right ),
\label{1203}\\
F_2(\Delta)\equiv\Omega^{ren}_{\scriptscriptstyle{T}}(M,\Delta)\Big
|_{M=0}\!
&=&\omega_2(\Delta)-2T\int_{-\infty}^{\infty}\frac{d^2p}{(2\pi)^2}\ln\left
(\big [1+e^{-\beta E^+_{\Delta}}\big ]\big [1+e^{-\beta
E^-_{\Delta}}\big ]\right ),\label{1204}
\end{eqnarray}
where $E=\sqrt{|\vec p|^2+M^2}$, $(E^+_{\Delta})^2=(|\vec
p|\pm\mu)^2+\Delta^2$, and the functions $\omega_{1}(M)$,
$\omega_{2}(\Delta)$ are presented in (\ref{26}) and (\ref{27}),
respectively. The gaps $M_0$ and $\Delta_0$ are the
solutions of the following stationary (gap) equations,
\begin{eqnarray}
&&\frac{\partial F_1(M)}{\partial M}\equiv\frac{M}{\pi}f_1(M)=0,~~~
\frac{\partial F_2(\Delta)}{\partial \Delta}
\equiv\frac{\Delta}{\pi}f_2(\Delta)=0,\label{31}
\end{eqnarray}
where
\begin{eqnarray}
&&f_1(M) =\frac 1{g_1}
+M+T\ln\left\{ \big [1+e^{-\beta (M+\mu)}\big ]\big [1+e^{-\beta
(M-\mu)}\big ]\right\},\label{310}\\
%\end{eqnarray}
%\begin{eqnarray}
f_2(\Delta)
&=&\frac
1{g_2}+\sqrt{\mu^2+\Delta^2}+2T\ln\left
(1+e^{-\beta\sqrt{\mu^2+\Delta^2}}\right )-\mu\int_0^\mu\tanh \Big
(\frac{\beta\sqrt{q^2+\Delta^2}}{2}\Big
)\frac{dq}{\sqrt{q^2+\Delta^2}}, \label{32}
\end{eqnarray}
respectively (for details, see Appendix \ref{ApC}). On the basis of
these gap equations we will study the phase structure  of the model
at $T>0$.

{\bf The case $g_1>0$.} First of all let us consider the phase
portrait  of the model at $g_1>0$. It straightforwardly follows from
(\ref{31}) and (\ref{310}) that the gap $M_0$ is always zero at
$g_1>0$ (it is a nonzero quantity only at  $g_1<0$). However, the
gap $\Delta_0$ is positive both at $g_2<0$ and $g_2>0$ if
temperature is sufficiently small, i.e. when $f_2(0)<0$. So, at
$g_1>0$ and for arbitrary values of $\mu$ the superconducting phase
III is arranged at sufficiently small values of temperature
$T<T_c(\mu)$. At $T>T_c(\mu)$ the gap equations (\ref{31}) supply
the $\Delta_0=0$ and $M_0=0$ gap values, i.e. the symmetric phase. The
second order phase transition temperature $T_c(\mu)$ is the solution
of the equation $f_2(0)=0$,
\begin{eqnarray}
f_2(0)\equiv\frac 1{g_2}+\mu+2T\ln\left
(1+e^{-\beta\mu}\right )-\mu\int_0^\mu\tanh \Big
(\frac{\beta q}{2}\Big
)\frac{dq}{q}=0. \label{33}
\end{eqnarray}
Hence, in the $(\mu,T)$-plane the curve $T=T_c(\mu)$ is the boundary
between symmetric and superconducting phases. Numerical
investigation of the equation (\ref{33}) produces at $g_2=\pm
0.5g_1$ the phase portraits of the model presented in Figs 11 and
12.

If $g_2<0$, then it follows from (\ref{33}) that $T_c(0)=-1/(2g_2ln
2)$. Moreover,  in this case the critical temperature can be given
as a series over the small parameter $\mu$,
\begin{eqnarray}
T_c(\mu)=T_c(0)-\mu^2g_2/16+o(\mu^2g_2). \label{34}
\end{eqnarray}
Comparing this expansion at $g_2=-0.5g_1$ with $T_c(\mu)$ of Fig.
11, we see  that (\ref{34}) supplies a rather good approximation for
the critical temperature only in the interval $0<\mu g_1<0.2$.

Now let us try to present some analytical approximation for the
$T_c(\mu)$  at $g_2>0$ ($g_1$ is still fixed and positive). For this
purpose note first of all that for all points of the critical curve
$T=T_c(\mu)$ of Fig. 12 the relation $\mu /T\equiv\mu\beta >>1$ is
valid.
\begin{figure}
%----figure 13,14
 \includegraphics[width=0.45\textwidth]{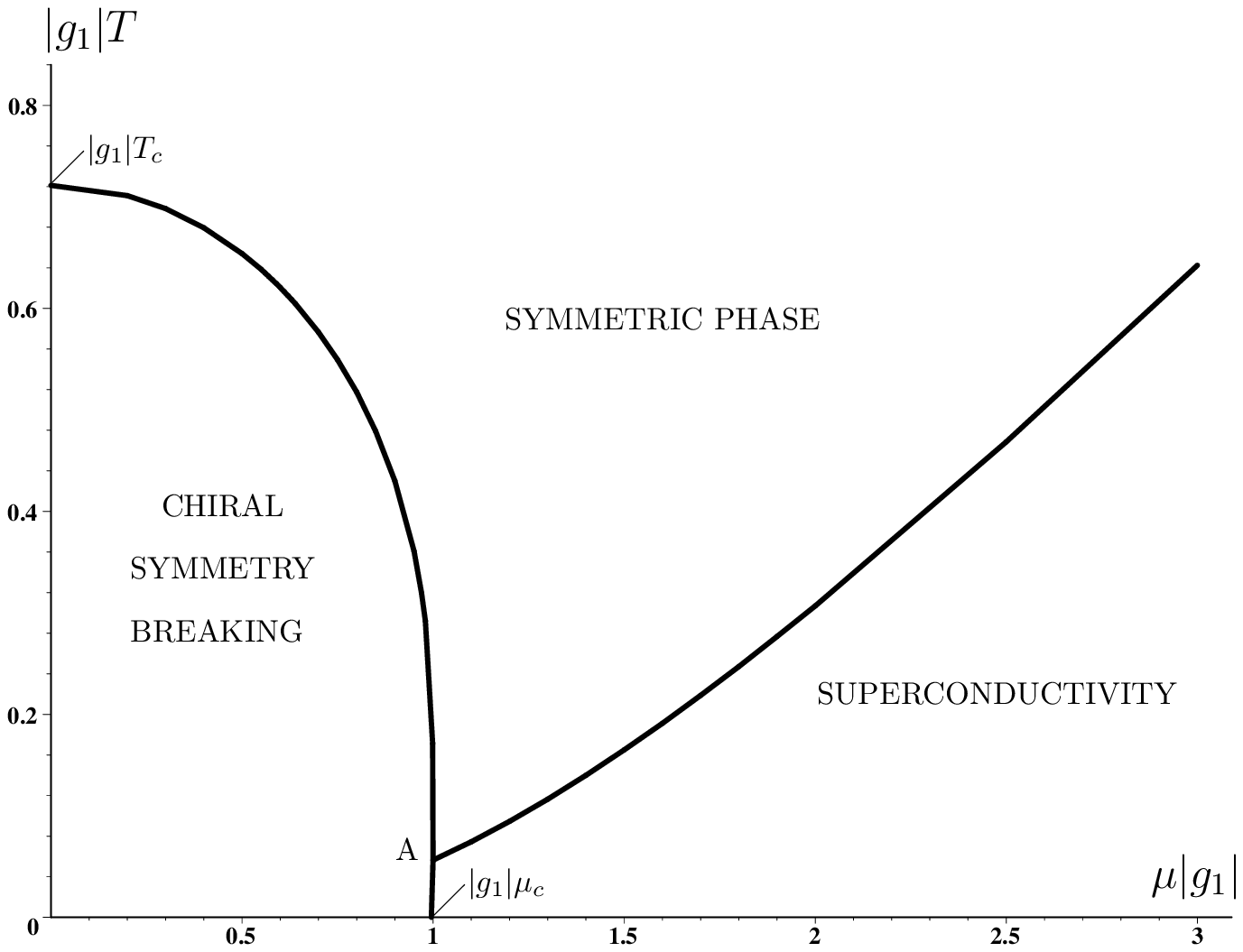}
 \hfill
 \includegraphics[width=0.45\textwidth]{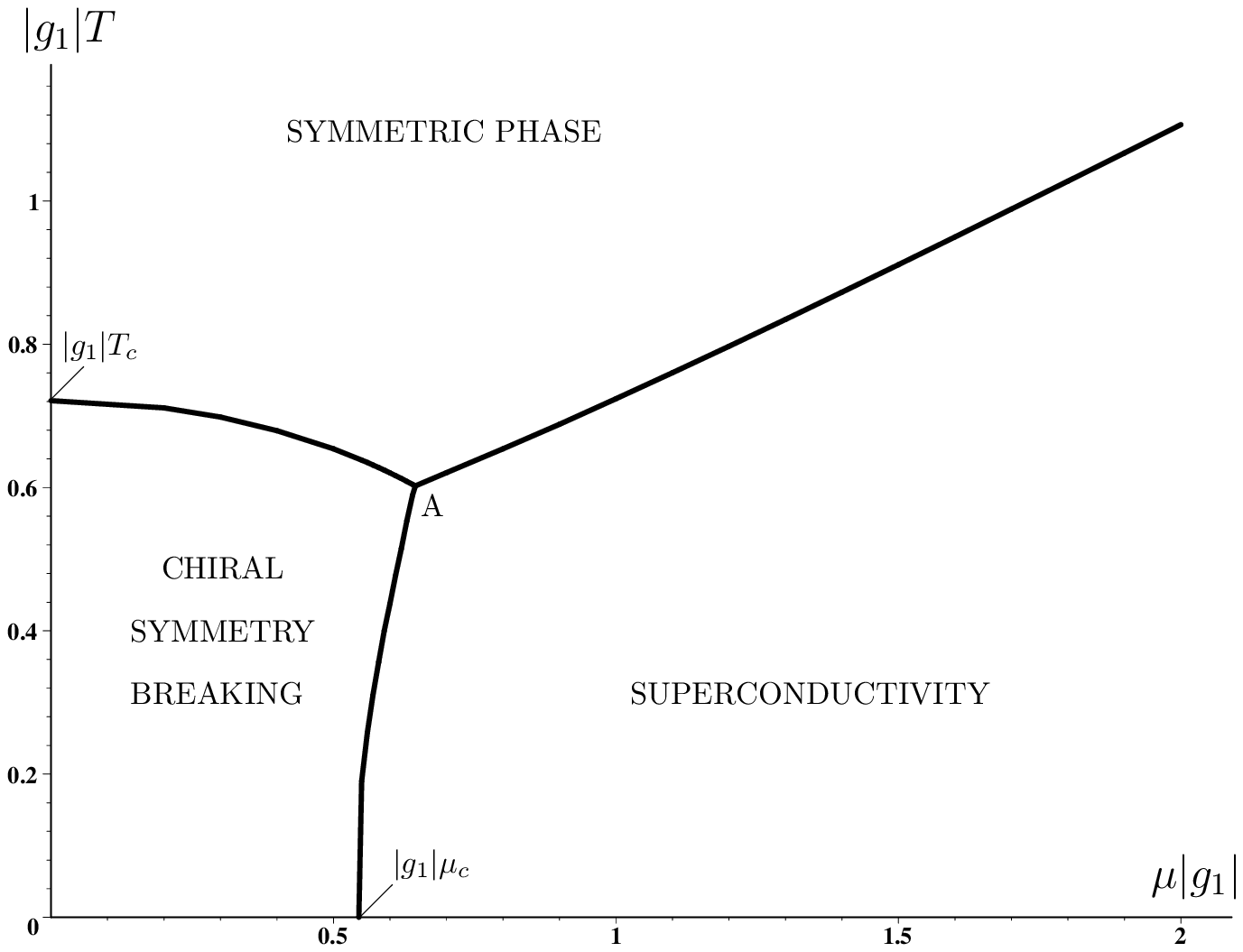}\\
\parbox[t]{0.45\textwidth}{
 \caption{$(\mu,T)$-phase diagram of the model at
$g_2=0.5|g_1|$ and arbitrary fixed $g_1<0$. All the curves are the
lines of second order phase transitions except the boundary between
the superconducting and chiral symmetry breaking phases, where a
first order phase transition is realized. The coordinates of the
tricritical point A are the following ones, $|g_1|\mu_A\approx
0.999$ and $|g_1|T_A\approx 0.056$. Moreover,
$|g_1|\mu_c\approx 0.995$ and $|g_1|T_c=1/(2\ln2)\approx 0.721$. }
 }\hfill
\parbox[t]{0.45\textwidth}{
\caption{$(\mu,T)$-phase diagram of the model at
$g_2=-1.5|g_1|$ and arbitrary fixed $g_1<0$. All the curves are the
lines of second order phase transitions except the boundary between
the superconducting and chiral symmetry breaking phases, where a
first order phase transition is realized. The coordinates of the
tricritical point A are the following ones, $|g_1|\mu_A\approx
0.645$ and $|g_1|T_A\approx 0.602$. Moreover,
$|g_1|\mu_c\approx 0.545$ and $|g_1|T_c=1/(2\ln2)\approx 0.721$. }  }
\end{figure}
Then, it is convenient to present the equation (\ref{33}) in the
following equivalent form:
\begin{eqnarray}
\frac 1{2Tg_2}+\frac{\mu\beta}2+\ln\left (1+e^{-\beta\mu}\right
)-\frac{\mu\beta}2\left\{C_1+\int_1^{\mu\beta/2}
\frac{dz}{z}+C_2-\int^\infty_{\mu\beta/2}[\tanh
z-1]\frac{dz}{z}\right\}=0, \label{35}
\end{eqnarray}
where
\begin{eqnarray}
C_1=\int^1_{0}\tanh z\frac{dz}{z}\approx 0.910,~~~~~
C_2=\int^\infty_{1}[\tanh z-1]\frac{dz}{z}\approx -0.091.
\label{36}
\end{eqnarray}
The third term in (\ref{35}) as well as the last integral in the braces of
(\ref{35}) can be neglected in comparison with other terms. The
obtained equation can be easily solved with respect to T. As a
result we have
\begin{eqnarray}
T_c(\mu)\approx\frac {\mu}2\exp \left [C_1+C_2-1-1/(\mu g_2)\right ].
\label{37}
\end{eqnarray}
Note that at $g_2=0.5g_1$ the plot of the expression (\ref{37})
coincides with great accuracy with the critical temperature of Fig.
12 in the whole interval $0<\mu g_1<2$.

{\bf The case $g_1<0$.} In this case we present three
$(\mu,T)$-phase  portraits of the model for qualitatively distinct
relations between $g_1$ and $g_2$. The first one for $g_2=-0.5|g_1|$
(which is in Fig. 11) was already described above because it is the
same as in the case $g_1>0$, $g_2=-0.5g_1$. The other two phase
portraits are represented in Figs. 13 and 14 for $g_2=0.5|g_1|$ and
$g_2=-1.5|g_1|$, respectively. There the points $(\mu,T)$ of the
boundary between the symmetric and chiral symmetry breaking (or
superconducting) phases  are given implicitly by the equation
$f_1(0)=0$ (or $f_2(0)=0$), where the functions $f_1(M)$ and
$f_2(\Delta)$ are defined in (\ref{310}) and  (\ref{32}),
respectively. On these boundaries, the second order phase
transitions occur. In contrast, the boundary between chiral symmetry
breaking and superconducting phases is the curve of the first order
phase transitions. So at the points $(\mu,T)$ of this boundary the
two phases may coexist.

Analyzing the cited above $(\mu,T)$-phase diagrams of Figs. 11--14,
we see that for each arbitrary fixed value $T$ of the temperature
(and for all relations between coupling constants) there exist a
definite value $\mu_T$ of the chemical potential such that for all
$\mu>\mu_T$ the superconducting phase is realized in the system.
This property is inherent only to a (2+1)-dimensional model (1) and
it is absent in the two-dimensional analogue \cite{chodos}.

\section{Summary and conclusions}

In this paper we study the competition between chiral and
superconducting condensations in the framework of the
(2+1)-dimensional GN-type model (1) which is a direct generalization
of the two-dimensional analogue by Chodos et al. \cite{chodos}. So,
the initial four-fermion model (1) describes interactions both in
the fermion-antifermion (or chiral) and superconducting difermion
(or Cooper pairing) channels with couplings $G_1$ and $G_2$,
respectively. Moreover, it is chirally and $U(1)$ invariant one
(the last group corresponds to conservation of the fermion number
or electric charge of the system). To avoid the ban on the
spontaneous breaking of continuous symmetry in (2+1)-dimensional
field theories at $T>0$, we consider, as it was done in
\cite{chodos}, the phase structure of our model in the leading order
of the large-$N$ technique, i.e. in the limit $N\to\infty$, where
$N$ is a number of fermion fields.

{\bf The case $T=0$, $\mu=0$.} First of all we have investigated the
thermodynamic potential of the model at $T=0$, $\mu=0$. In
this case the phase portrait is presented in Fig. 1 in terms of the
renormalization group invariant finite coupling constants $g_1$ and
$g_2$. Each point $(g_1,g_2)$ of this diagram corresponds to a
definite phase. For example, at $g_{1,2}>0$, i.e. at sufficiently
small values of the bare coupling constants $G_{1,2}$ (see the
comment at the end of Section \ref{mu0}), neither chiral nor $U(1)$
symmetries are violated and the system is in the symmetric
phase, etc. %Moreover, it turns out that the two phases, chiral
%symmetry breaking II and superconducting III, may coexist in this
%diagram only on the line $g_1=g_2\equiv g$, with $g<0$.

{\bf The case $T=0$, $\mu\ne 0$.} In this case we select two
qualitatively different situations, $g_1<0$ and $g_1>0$. If $g_1<0$
and fixed, then in Fig. 2 we draw the $(g_2,\mu)$-phase diagram of
the model. It means that at $g_2>0$ or at $g_2<g_1$ the phase II
with zero particle density is realized at sufficiently low values of
$\mu$. In this case the ground state of the system is an empty
space. Then at some critical value $\mu=\mu_{crit}(g_2)$ bubbles of
the new phase III with particle density $n_{crit}(g_2)$ (see Fig. 4) can appear in the space, and for all
$\mu>\mu_{crit}(g_2)$ the whole space is filled with superconducting
phase, in which particle density $n$ is not zero, $n>n_{crit}(g_2)$. If $g_1>0$, then the
system is in the superconducting phase even at arbitrary small
values of $\mu$. Hence, at $T=0$ and at growing chemical potential
the system is transformed into a superconducting state.

{\bf The case $T>0$, $\mu\ne 0$.} Phase portraits of the model are
presented in this case in Figs 11--14. It is clear from the figures
that at fixed $\mu$ and increasing temperature the symmetric phase
is restored. However, at arbitrary fixed $T$,  growth of the
chemical potential leads to appearing of superconductivity
in the system at arbitrary relations between coupling constants $g_1$
and $g_2$.

The fact that chemical potential induces superconductivity phenomenon
is the main result of our paper. Note that in general this property of
the (2+1)-dimensional GN-type model (1) is not valid in the case of
the two-dimensional model \cite{chodos}.

We hope that our investigations can shed new light on the
superconducting phenomena in condensed matter systems with planar
structures.

\appendix
\section{Algebra of the $\gamma$-matrices in the case of SO(2,1) group}
\label{ApA}

The two-dimensional irreducible representation of the 3-dimensional
Lorentz group SO(2,1) is realized by the following $2\times 2$
$\tilde\gamma$-matrices:
\begin{eqnarray}
\tilde\gamma^0=\sigma_3=
\left (\begin{array}{cc}
1 & 0\\
0 &-1
\end{array}\right ),\,\,
\tilde\gamma^1=i\sigma_1=
\left (\begin{array}{cc}
0 & i\\
i &0
\end{array}\right ),\,\,
\tilde\gamma^2=i\sigma_2=
\left (\begin{array}{cc}
0 & 1\\
-1 &0
\end{array}\right ),
\label{A1}
\end{eqnarray}
acting on two-component Dirac spinors.

They have the properties:
\begin{eqnarray}
Tr(\tilde\gamma^{\mu}\tilde\gamma^{\nu})=2g^{\mu\nu};~~
[\tilde\gamma^{\mu},\tilde\gamma^{\nu}]=-2i\varepsilon^{\mu\nu\alpha}
\tilde\gamma_{\alpha};~
~\tilde\gamma^{\mu}\tilde\gamma^{\nu}=-i\varepsilon^{\mu\nu\alpha}
\tilde
\gamma_{\alpha}+
g^{\mu\nu},
\label{A2}
\end{eqnarray}
where $g^{\mu\nu}=g_{\mu\nu}=diag(1,-1,-1),
~\tilde\gamma_{\alpha}=g_{\alpha\beta}\tilde\gamma^{\beta},~
\varepsilon^{012}=1$.
There is also the relation:
\begin{eqnarray}
Tr(\tilde\gamma^{\mu}\tilde\gamma^{\nu}\tilde\gamma^{\alpha})=
-2i\varepsilon^{\mu\nu\alpha}.
\label{A3}
\end{eqnarray}
Note that the definition of chiral symmetry is slightly unusual in
three dimensions (spin is here a pseudoscalar rather than a (axial)
vector). The formal reason is simply that there exists no other $2\times 2$
matrix anticommuting with the Dirac matrices $\tilde\gamma^{\nu}$
which would allow the introduction of a $\gamma^5$-matrix in the
irreducible representation. The important concept of 'chiral'
symmetries  and their breakdown by mass terms can nevertheless be
realized also in the framework of (2+1)-dimensional quantum field
theories by considering a four-component reducible representation
for Dirac fields. In this case the Dirac spinors $\psi$ have the
following form:
\begin{eqnarray}
\psi(x)=
\left (\begin{array}{cc}
\tilde\psi_{1}(x)\\
\tilde\psi_{2}(x)
\end{array}\right ),
\label{A4}
\end{eqnarray}
with $\tilde\psi_1,\tilde\psi_2$ being two-component spinors.
In the reducible four-dimensional spinor representation one deals
with
(4$\times$4) $\gamma$-matrices:
$\gamma^\mu=diag(\tilde\gamma^\mu,-\tilde\gamma^\mu)$, where
$\tilde\gamma^\mu$ are given in (\ref{A1}). One can easily show, that
($\mu,\nu=0,1,2$):
\begin{eqnarray}
&&Tr(\gamma^\mu\gamma^\nu)=4g^{\mu\nu};~~
\gamma^\mu\gamma^\nu=\sigma^{\mu\nu}+g^{\mu\nu};~~\nonumber\\
&&\sigma^{\mu\nu}=\frac{1}{2}[\gamma^\mu,\gamma^\nu]
=diag(-i\varepsilon^{\mu\nu\alpha}\tilde\gamma_\alpha,
-i\varepsilon^{\mu\nu\alpha}\tilde\gamma_\alpha).
\label{A5}
\end{eqnarray}
In addition to the  Dirac matrices $\gamma^\mu~~(\mu=0,1,2)$ there
exist two other matrices $\gamma^3$, $\gamma^5$ which anticommute
with all $\gamma^\mu~~(\mu=0,1,2)$ and with themselves
\begin{eqnarray}
\gamma^3=
\left (\begin{array}{cc}
0~,& I\\
I~,& 0
\end{array}\right ),\,
\gamma^5=\gamma^0\gamma^1\gamma^2\gamma^3=
i\left (\begin{array}{cc}
0~,& -I\\
I~,& 0
\end{array}\right ),\,\,
\label{A6}
\end{eqnarray}
with  $I$ being the unit $2\times 2$ matrix. Finally note that in
terms  of two-component spinors $\tilde\psi_1,\tilde\psi_2$ the
parity transformation $P$, defined in the space of four-component
spinors by the relation (\ref{03}), looks like
\begin{eqnarray}
P:~~~\tilde\psi_1 (t,x,y)\to i\tilde\gamma^1\tilde\psi_2 (t,-x,y);~~ \tilde\psi_2 (t,x,y)\to i\tilde\gamma^1\tilde\psi_1 (t,-x,y).
\label{A7}
\end{eqnarray}
Such a definition of the space parity transformation is commonly used
in (2+1)-dimensional theories with four-component representation for
Dirac spinors (see, e.g., in \cite{appelquist}).

\section{The path integration over anticommutating fields}
\label{ApB}

Let us calculate the following path integral over anticommutating
four-component Dirac spinor fields $q(x)$, $\bar q(x)$:
\begin{eqnarray}
I=\int[d\bar q][dq]\exp\Big (i\int d^3 x\Big [\bar q  D
q-\frac{\Delta}{2}(q^TC q) -\frac{\Delta}{2}(\bar q C\bar q^T)\Big ]
\Big ) \label{B1},
\end{eqnarray}
where we use the notations of Section \ref{effaction} and, in
particular, the operator $D$ is given in (\ref{9}). Note in
addition, the integral $I$ is equal to the argument of the
$\ln$-function in the formula (\ref{9}) in the particular case
$N=1$. Recall, there are general Gaussian path integrals
\cite{vasiliev}:
\begin{eqnarray}
\int[dq]\exp\Big (i\int d^3 x\Big
[-\frac{1}{2}q^T A q +\eta^Tq
\Big ]\Big )&=&\left(\det(A)\right )^{1/2}\exp\Big (-\frac{i}{2}\int d^3 x\Big [\eta^T A^{-1}\eta\Big ]\Big ),  \label{B2}\\
\int[d\bar q]\exp\Big (i\int d^3 x\Big
[-\frac{1}{2}\bar q A \bar q^T +\bar\eta\bar q^T\Big ] \Big )&
=&\left (\det(A)\right )^{1/2}\exp\Big (-\frac{i}{2}\int d^3 x\Big
[\bar\eta A^{-1}\bar\eta^T\Big ]\Big ), \label{B3}
\end{eqnarray}
where $A$ is an antisymmetric operator in coordinate and spinor
spaces, and $\eta(x)$, $\bar\eta (x)$ are anticommutating spinor
sources which also anticommutate with $q$ and $\bar q$. First, let us
integrate in (\ref{B1}) over $q$-fields with the help of the
relation (\ref{B2}) supposing there that $A=\Delta C$, $\bar
qD=\eta^T$, i.e. $\eta=D^T\bar q^T$. Then
\begin{eqnarray}
I=\left (\det(\Delta C)\right )^{1/2}\int[d\bar q]\exp\Big
(-\frac{i}{2}\int d^3 x \bar q \big [\Delta C+D(\Delta
C)^{-1}D^T\big ]\bar q^T\Big ). \label{B4}
\end{eqnarray}
Second, the integration over $\bar q$-fields in (\ref{B4}) can be
easily performed with the help of the formula (\ref{B3}), where one
should put $A=\Delta C+D(\Delta C)^{-1}D^T$ and $\bar\eta=0$. As a
result, we have
\begin{eqnarray}
I=\left (\det(\Delta C)\right )^{1/2}\left (\det[\Delta C+D(\Delta
C)^{-1}D^T]\right )^{1/2}=\left (\det [\Delta^2C^2+D C^{-1} D^T
C]\right )^{1/2}. \label{B5}
\end{eqnarray}
Taking into account the relations $(\partial_\nu)^T=-\partial_\nu$
and $C^{-1} (\gamma^\nu)^T C =-\gamma^\nu$ ($\nu=0,1,2$), we obtain
from (\ref{B5})
\begin{eqnarray}
I=\left (\det [-\Delta^2+D_+D_-]\right )^{1/2}\equiv\left (\det
B\right )^{1/2}, \label{B7}
\end{eqnarray}
where $D_\pm=\gamma^\nu i\partial_\nu-M\pm \mu\gamma^0$. Using the
general relation $\det B =\exp ({\rm Tr}\ln B)$, we get from
(\ref{B7}):
\begin{eqnarray}
\ln I=\frac 12 {\rm Tr}\ln\left (B\right
)=\sum_{i=1}^{2}\int\frac{d^3p}{(2\pi)^3} \ln(\lambda_i(p))\int
d^3x. \label{B8}
\end{eqnarray}
(A more detailed consideration of operator traces is presented in
Appendix A of the paper \cite{Ebert:2009ty}.) In this formula symbol
Tr means the trace of an operator both in the coordinate and
internal spaces. Moreover, $\lambda_i(p)$ ($i=1,2$) in (\ref{B8})
are two twice degenerated eigenvalues of the 4$\times$4 Fourier
transformation matrix $\bar B(p)$ of the operator $B$, i.e.
\begin{eqnarray}
\lambda_{1,2}(p)&=&M^2-p_1^2-p_2^2-\mu^2+p_0^2-\Delta^2 \pm
2\sqrt{-M^2p_2^2-M^2p_1^2+M^2p_0^2+\mu^2p_2^2+\mu^2p_1^2}.
\label{B11}
\end{eqnarray}

\section{Gap equations}
\label{ApC}

The equation for the gap $M_0$, i.e. the first one of equations
(\ref{31}), is obtained, e.g., in \cite{klimenko2}, where a phase
structure of the initial model (1) was consided in the particular case
of $G_2=0$.

To obtain a gap equation for the superconducting gap $\Delta_0$,
$\partial F_2(\Delta)/\partial \Delta=0$, let us first transform the
original expression (\ref{1204}) for the TDP $F_2(\Delta)$ using polar
coordinates in the integral in (\ref{1204}). Integrating in the
obtained expression over a polar angle, we have
\begin{eqnarray}
F_2(\Delta)=\omega_2(\Delta)-\frac{T}{\pi}\int_0^{\infty}pdp\ln\left
(1+e^{-\beta \sqrt{(p+\mu)^2+\Delta^2}}\right )-\frac{T}{
\pi}\int_0^{\infty}pdp\ln\left
(1+e^{-\beta \sqrt{(p-\mu)^2+\Delta^2}}\right ).\label{124}
\end{eqnarray}
It is very convenient to change integration variables in (\ref{124}) (we use
$q=p+\mu$ for the first integral and $q=p-\mu$ for the second one) and,
after some manipulations, to get an equivalent expression,
\begin{eqnarray}
F_2(\Delta)=\omega_2(\Delta)-\frac{2T}{\pi}\int_\mu^{\infty}qdq\ln\left
(1+e^{-\beta \sqrt{q^2+\Delta^2}}\right )-\frac{2T\mu}{
\pi}\int_0^{\mu}dq\ln\left
(1+e^{-\beta \sqrt{q^2+\Delta^2}}\right ).\label{1240}
\end{eqnarray}
Starting from (\ref{1240}) and taking into account the expression
(\ref{27}) for $\omega_2(\Delta)$, we have the following gap equation:
\begin{eqnarray}
\frac{\partial F_2(\Delta)}{\partial \Delta}&=&\frac{\Delta}{\pi
g_2}+\frac{\Delta}{\pi}\sqrt{\mu^2+\Delta^2}-\frac{\mu\Delta}{\pi}\ln\left
(\frac{\mu+\sqrt{\mu^2+\Delta^2}}{\Delta}\right )\nonumber\\
&+&\frac{\Delta}{\pi}\int_\mu^\infty\frac{2qdq}{\sqrt{q^2+\Delta^2}\left
(1+e^{\beta\sqrt{q^2+\Delta^2}}\right
)}+\frac{2\Delta\mu}{\pi}\int_0^\mu\frac{dq}{\sqrt{q^2+\Delta^2}\left
(1+e^{\beta\sqrt{q^2+\Delta^2}}\right )}=0.\label{C2}
\end{eqnarray}
The first integral in (\ref{C2}) is a rather simple one, i.e.
\begin{eqnarray}
\int_\mu^\infty\frac{2qdq}{\sqrt{q^2+\Delta^2}\left
(1+e^{\beta\sqrt{q^2+\Delta^2}}\right )}=\frac{2}{\beta}\ln\left
(1+e^{-\beta\sqrt{\mu^2+\Delta^2}}\right ).\label{C3}
\end{eqnarray}
In contrast, let us present the third term in (\ref{C2}) in the
integral form, i.e.
\begin{eqnarray}
-\frac{\mu\Delta}{\pi}\ln\left
(\frac{\mu+\sqrt{\mu^2+\Delta^2}}{\Delta}\right
)=-\frac{\mu\Delta}{\pi}\int_0^\mu\frac{dq}{\sqrt{q^2+\Delta^2}},\label{C4}
\end{eqnarray}
which then can be combined with the last integral of (\ref{C2}). As
a result we obtain for the superconducting gap $\Delta_0$ the
second of equations (\ref{31}).

\end{document}